\documentclass[a4paper,showpacs,twocolumn,amsmath,amssymb,superscriptaddress ,floatfix,prb,preprintnumbers,footinbib]{revtex4-1}

\usepackage[utf8]{inputenc}
\usepackage{graphicx}
\usepackage{subfigure}
\usepackage[colorlinks=true,citecolor=blue]{hyperref}
\usepackage{bm}

\begin{document}
\newcommand{\dauf}[1]{ d^{\dag}_{ #1 }}
\newcommand{\dab}[1]{d^{\phantom{\dag}}_{ #1 }}
\newcommand{\cauf}[1]{ c^{\dag}_{ #1 }}
\newcommand{\cab}[1]{ c^{\phantom{\dag}}_{ #1 }}
\newcommand{\s}{\sigma}
\newcommand{\e}{\epsilon}
\newcommand{\la}{\lambda}
\newcommand{\n}{\alpha}
\newcommand{\w}{\omega}
\newcommand{\K}{\bm{\mathcal{W}}}
\newcommand{\p}{^\prime}
\newcommand{\up}{\uparrow}
\newcommand{\down}{\downarrow}
\newcommand{\de}{\Delta E}
\newcommand{\mde}{\bm{\Delta\mathcal{E}}}
\newcommand{\dep}{\Delta \epsilon}
\newcommand{\bdep}{\overline{\Delta \epsilon}}
\newcommand{\be}{\overline{\epsilon}}
\newcommand{\aso}{\alpha_{\text{so}}}
\newcommand{\ddt}{\: \frac{\mathrm{d}}{\mathrm{d}t}}
\newcommand{\dw}{\: \mathrm{d}\omega \: }
\newcommand{\dt}{\: \mathrm{d}t \: }
\newcommand{\dtone}{\: \mathrm{d}t_1 \: }
\newcommand{\dttwo}{\: \mathrm{d}t_2 \: }
\newcommand{\dtp}{\: \mathrm{d}t^\prime \: }
\newcommand{\dtpp}{\: \mathrm{d}t^{\prime\prime} \: }
\newcommand{\trace}[1]{\: \mathrm{Tr}\! \left[ #1  \right]}
\newcommand{\bra}[1]{\left\langle #1\right|}
\newcommand{\ket}[1]{\left|#1\right\rangle}
\newcommand{\etal}{~\textit{et~al.}}
\newcommand{\ci}{Coulomb interaction}
\newcommand{\so}{SO}

\title{Adiabatic pumping through an interacting quantum dot with spin-orbit coupling}

\author{Stephan Rojek} 
\affiliation{Theoretische Physik, Universit\"at Duisburg-Essen and CENIDE, 47048 Duisburg, Germany}
\author{J\"urgen K\"onig} 
\affiliation{Theoretische Physik, Universit\"at Duisburg-Essen and CENIDE, 47048 Duisburg, Germany}
\author{Alexander Shnirman}
 \affiliation{Institut f\"ur Theorie der Kondensierten Materie and DFG Center for Functional Nanostructures (CFN), Karlsruhe Institute of Technology, 76128 Karlsruhe, Germany}

\date{\today}

\begin{abstract}
We study adiabatic pumping through a two-level quantum dot with spin-orbit coupling.
Using a diagrammatic real-time approach, we calculate both the pumped charge and spin for a periodic variation of the dot's energy levels in the limit of weak tunnel coupling.
Thereby, we compare the two limits of vanishing and infinitely large charging energy on the quantum dot.
We discuss the dependence of the pumped charge and pumped spin on gate voltages, the symmetry in the tunnel-matrix elements and spin-orbit coupling strength.
We identify the possibility to generate pure spin currents in the absence of charge currents.
\end{abstract}






\maketitle
\section{Introduction}

A central issue in the field of spintronics is the design of spin-based electronic devices.\cite{wolf_spintronics:spin-based_2001,awschalom_challenges_2007}  
They may involve ferromagnets or external magnetic fields to control the spin degree of freedom.\cite{spintronics_2004} 
But recently, all-electric spintronic devices also have gained interest.\cite{awschalom_spintronics_2009} 
They rely on spin-orbit (\so) interaction, the strength of which is tunable via external gates in semiconductor heterostructures,\cite{nitta_gate_1997,papadakis_effect_1999} a basic requirement for the realization of a spin field-effect transistor.\cite{*[{}] [{ [JETP Lett. \textbf{39}, 78 (1984)].}] Bychkov1984, bychkov_oscillatory_1984, datta_electronic_1990,koo_control_2009}

A spin-polarized current in a semiconductor can be generated by spin injection.\cite{*[{}] [{ [Sov. Phys. Semicond., Vol. \textbf{10}, 698 (1976)].}] Aronov1976,Hammar1999,monzon_spin_1999,Filip2000,zhu_room-temperature_2001}
Here we focus on an alternative route that relies on pumping. 
By varying the parameters of a mesoscopic system periodically in time, a finite charge or spin current can be sustained. 
Experimental studies have investigated charge pumping in several mesoscopic devices.\cite{pothier_single-electron_1992,switkes_adiabatic_1999,fuhrer_single_2007,buitelaar_adiabatic_2008,kaestner_robust_2008} 
Spin pumping has been experimentally realized in the presence of an external magnetic field.\cite{watson_experimental_2003} 
Theoretical studies of spin pumping involve external magnetic fields,\cite{mucciolo_adiabatic_2002} ferromagnetic leads,\cite{wu_spin-polarized_2002,splettstoesser_adiabatic_2008,riwar_charge_2010} and also \so~coupling.\cite{sharma_mesoscopic_2003,governale_pumping_2003,brosco_prediction_2010}

 In the present paper, we consider the minimal model that contains \so~interaction: a quantum dot with two spin-degenerate orbital levels. Such a two-level quantum dot with more than two leads has been suggested as a spin filter.\cite{eto_quantum_2010}
 We focus on the \emph{adiabatic} limit of pumping, i.e., the parameters are varied slowly in time compared to the dwell time of the mesoscopic system.\cite{thouless_quantization_1983}
Adiabatic pumping of charge and spin through such a two-level dot has been considered in the limit of vanishing charging energy.\cite{brosco_prediction_2010} 
It was found that this system can act as an \textit{all-electric spin battery}, i.e., a finite spin current can be achieved without ferromagnets by electrically controlling the dot parameters.
For specific symmetries in the tunnel coupling of the dot to the leads even pure spin currents have been suggested.
From the analysis of Ref.~\onlinecite{brosco_prediction_2010}, which was based on a scattering-matrix approach,\cite{bttiker_current_1994,brouwer_scattering_1998,avron_geometry_2000} it is not clear whether and how the conclusions can be transferred to quantum dots with non vanishing \ci.
To answer this question is the main goal of the present paper.

In order to take the \ci~into account, we use a diagrammatic real-time approach\cite{schoeller_mesoscopic_1994,knig_resonant_1996,knig_zero-bias_1996} that allows for arbitrary strengths of the \ci.
We focus on the limit of weak tunnel coupling, for which we perform a systematic perturbation expansion to lowest order. 
To emphasize the role of \ci, we compare the limit of vanishing \ci~with the limit of an infinitely large charging energy.  
  
The paper is organized as follows. In Sec.~\ref{model} we introduce the model that describes the \so~interaction in a two-level quantum dot with \ci. Section~\ref{method} deals with the technique to calculate the pumped charge and pumped spin during one pumping cycle.
To study the dependence of the pumped charge (spin) on the four tunnel-matrix elements in a transparent way, we introduce in Sec.~\ref{isospin} an isospin representation of the orbital degree of freedom. 
Finally, in Sec.~\ref{res} we present the results for the pumped charge and pumped spin. 

\section{Model\label{model}}
\begin{figure}
	\includegraphics[width=.49\textwidth]{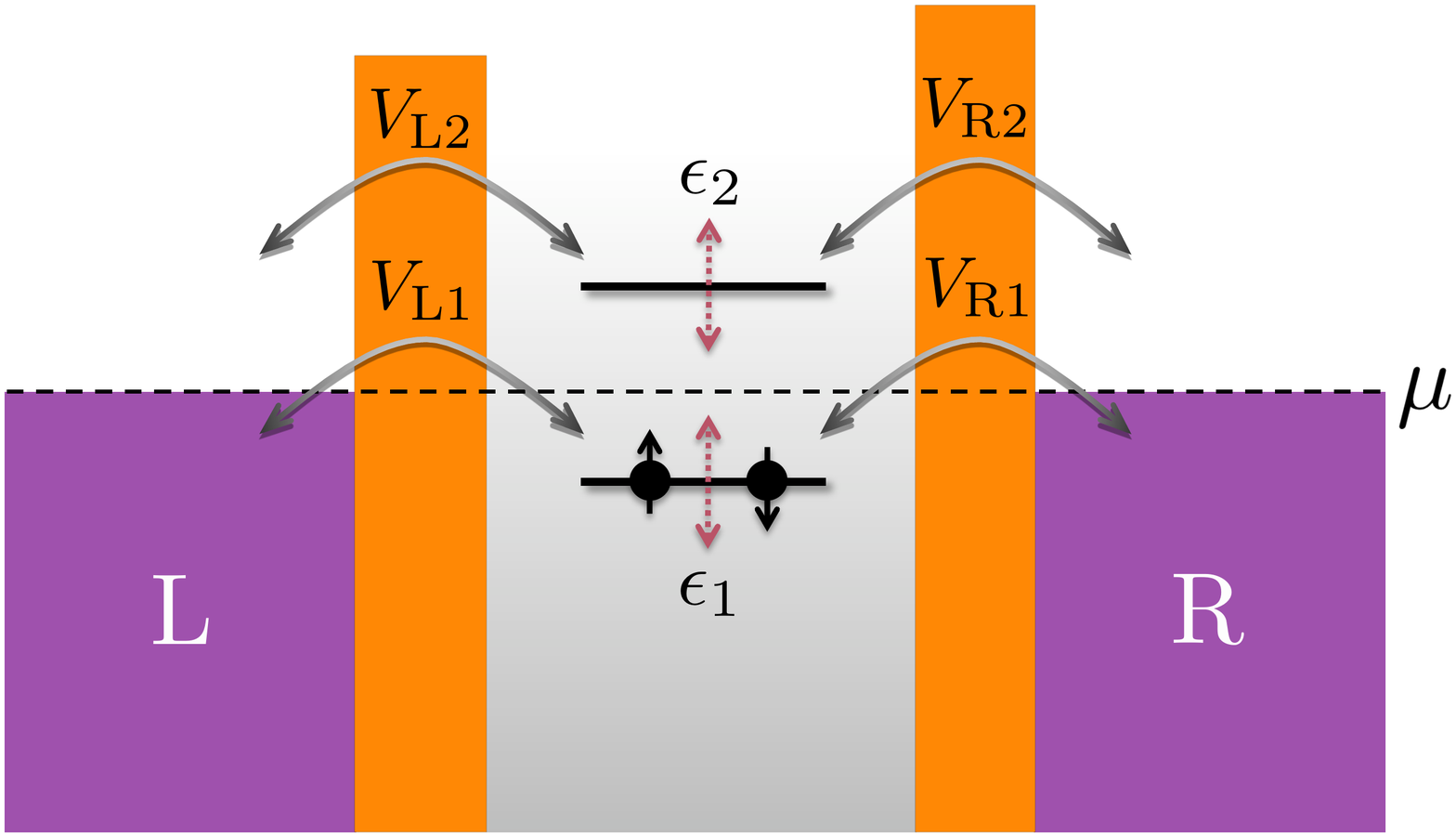}
	\caption{\label{fig:Model}(Color online) Energy scheme of the two-level quantum dot.
	The two orbital, spin-degenerate levels can be varied in time. They are tunnel coupled to the left (L) and the right (R) lead, with tunnel-matrix elements $V_{\la\n}$. The leads have the same chemical potential $\mu$.}
\end{figure}
We consider a quantum dot with two spin-degenerate orbital levels $\ket{\n\s}$ (with labels $\n=1,2$ for the orbital and $\s=\up,\down$ for the spin), tunnel coupled to the left (L) and the right (R) lead (see Fig.~\ref{fig:Model}). The system is described by the Hamiltonian
\begin{align}
 H &= H^\text{dot}  + H^{\text{lead}}+ H^{\text{tun}}\, .
\end{align} 
Here, $H^\text{dot}$ is the Hamiltonian of the isolated dot, $H^{\text{lead}}$ of the leads, and $H^{\text{tun}}$ of the tunneling between dot and leads. 

The Hamiltonian for the isolated quantum dot contains two parts.
The single-particle contribution for the two orbitals $\n$ with energy $\e_\n$, which are coupled by \so\ interaction, can be cast in the $4\times4$ matrix 
\begin{align}
	\begin{pmatrix}\e_1 \bm{\s}_0 & -i \bm{\aso} \cdot \bm{\s} \\i \bm{\aso} \cdot \bm{\s} & \e_2 \bm{\s} _0\end{pmatrix}\label{HDotMatrixGen} \, ,
\end{align} 
for the basis $\{ \ket{1\up}, \ket{1\down}, \ket{2\up}, \ket{2\down} \}$, where the spin quantization axis is chosen arbitrarily.
Here, $\bm{\s}$ denotes the vector of Pauli matrices, $\bm{\s}_0$ is the identity matrix, and $\bm{\aso}$ is a real vector describing the \so\ coupling.
The matrix in Eq.~\eqref{HDotMatrixGen} has the most general form that allows time-reversal symmetry. 
It has been used in the context of pumping\cite{brosco_prediction_2010} and was also recently applied to electron-transport in the presence of a magnetic field\cite{grap_interplay_2012} and to study the Josephson current through a double-dot structure.\cite{droste_josephson_2012}  
In the following, we choose the spin quantization axis parallel to $\bm{\aso}$ so the matrix becomes diagonal in spin space.

The second part of the dot Hamiltonian accounts for the charging energy $E_C (N-n_g)^2$, where $N$ is the total number of dot electrons and $n_g$ an external gate charge. 
Without loss of generality, we can choose $n_g=1/2$ (any other value can be achieved by a constant shift of the energies $\e_\n$).
This leads (up to an additive constant) to the dot Hamiltonian
\begin{align}
H^\text{dot} =& \sum_{\s\n} \e_\n\ \dauf{\n\s} \dab{\n\s} 
	+ \sum_{\s} i \s \aso \left( \dauf{2\s} \dab{1\s}-\text{ h.c.}\right)\nonumber\\
	&+ U \sum_{\n} n_{\n \up} n_{\n \down}  + U \sum_{\s\s\p} n_{1 \s} n_{2 \s\p}\, ,
\end{align} 
where the operator $\dauf{\n\s}$ creates an electron in state $\ket{\n\s}$ and the corresponding number operator is $n_{\n\s}=\dauf{\n\s}\dab{\n\s}$. We used the notation $\s=\pm1$ for spin parallel (antiparallel) to $\bm{\aso}$, 
$\aso=\left|\bm{\aso}\right|$, and $U=2E_C$.  

The leads are modeled as reservoirs of noninteracting electrons,
\begin{align}
	H^{\text{lead}} &= \sum_{\s k\la} \e_k \cauf{k \s \la} \cab{k \s \la} \, ,
\end{align}
where $\cauf{k \s \la}$ is the creation operator for an electron with spin $\s$ and momentum $k$ in lead $\la$.
Tunneling between dot and leads is described by the Hamiltonian
\begin{align} 
	H^{\text{tun}}&=\sum_{\s\n k\la} V_{\la \n} \cauf{k \s \la} \dab{\n \s} +\text{ h.c.}\, ,
\end{align}
with (spin-independent) tunnel-matrix elements $V_{\la \n}$ for tunneling between lead $\la$ and orbital $\n$. 

Pumping is achieved by varying system parameters periodically in time. 
In this paper, we assume that the energy levels $\e_\n(t)$ can be changed in time via external gates capacitively coupled to the system.
In principle, the external gates also may affect the \so~coupling, the tunnel couplings, and the electro chemical potential of the leads (via parasitic capacitances).
To simplify the discussion, however, we assume for the following these parameters to be constant in time.

We focus on the regime of adiabatic pumping, which is achieved for pumping frequencies $\Omega$ smaller than the inverse of the dwell time. This is valid for $\Omega \ll \Gamma$, where $\Gamma$ is the tunnel-coupling strength, $\Gamma=\sum_{\la\n}\Gamma_{\la \n \n}$, with $\Gamma_{\la \n \n\p}=2\pi\rho V_{\la\n\p}V^\ast_{\la\n}$. The density of states $\rho$ is assumed to be flat and equal for the left and right leads. We choose a gauge where all four tunnel-matrix elements are real.  

To study the effect of \ci, we compare results for the limit of noninteracting ($U=0$) and infinitely strong interacting ($U=\infty$) electrons on the dot.
In the latter case, the total number of electrons in the quantum dot can only be zero or 1.

\section{Method\label{method}}
To calculate the pumped charge and pumped spin, we use a diagrammatic real-time approach to adiabatic pumping through quantum-dot systems.~\cite{splettstoesser_adiabatic_2006}
For the present context, we extend the analysis of Ref.~\onlinecite{splettstoesser_adiabatic_2006} to allow for a time-dependent transformation of the basis states.
This is necessary since the \so~coupling couples time-dependent orbital levels, which, in turn, makes the dot eigenstates time dependent.

We start in Sec. \ref{density} with the kinetic equation for the reduced density matrix in its general form, which describes the time evolution of the dot's degrees of freedom.    
Subsequently, we perform both an adiabatic expansion, i.e., a perturbation expansion in the pumping frequency (Sec. \ref{adiabatic}) and a perturbation expansion in the tunnel-coupling strength (Sec. \ref{tunnel}) to describe the limit of weak tunnel coupling.
The pumped charge and pumped spin currents to lowest order in $\Gamma$ and $\Omega$ are derived in Sec. \ref{pumped}. 
Finally, in Sec. \ref{weak}, we perform the limit of weak pumping which assumes small amplitudes of the pumping parameters.

\subsection{Kinetic equation\label{density}}
The main idea of the diagrammatic real-time technique is based on the fact that the leads are described as large reservoirs of noninteracting electrons which can be integrated out in order to arrive at a reduced density matrix $\bm{p}$ for the dot degrees of freedom only.
For a matrix representation with matrix elements $p^{\chi_1}_{\chi_2}=\bra{\chi_1}\rho^{\text{dot}}\ket{\chi_2}$ (for the diagonal elements we introduce the notation $p_{\chi}\equiv p^{\chi}_{\chi}$), it is convenient to use the eigenstates $\ket{\chi_i}$ with corresponding eigenenergies $E_{\chi_i}$ as a basis.
For this, we employ a time-dependent unitary transformation $T$ acting on the dot Hamiltonian $H^\text{dot}$, such that $T^\dagger H^\text{dot} T$ is diagonal. 

The time evolution of the reduced density matrix is given by the kinetic equation
\begin{align}
\ddt \bm{p}(t)=&-\frac{i}{\hbar}\mde(t)\ \bm{p}(t)- \left[\bm{T}^\dagger\dot{\bm{T}}(t),\bm{p}(t)\right]\nonumber\\&+
\int^{t}_{-\infty}\dtp \K(t,t^\prime)\ \bm{p}(t^\prime)\, . \label{master}
\end{align}
The bold face indicates tensor notation.
The reduced density matrix $\bm{p}$ and $\bm{T}^\dagger\dot{\bm{T}}$ are tensors of rank 2, while $\mde$ and $\K$ are tensors of rank 4, i.e.,
\begin{align}
\left(\K(t,t\p)\bm{p}(t\p)\right)^{\chi_1}_{\chi_2}=\sum_{\chi^{\prime}_1,\chi^{\prime}_2}W^{\chi_1\chi_1^{\prime}}_{\chi_2\chi_2^{\prime}}(t,t\p)p^{\chi_1^{\prime}}_{\chi_2^{\prime}}(t\p)\, . \end{align} 
The kernel element $W^{\chi_1\chi_1^{\prime}}_{\chi_2\chi_2^{\prime}}(t,t\p)$ describes the transition from $p^{\chi_1^{\prime}}_{\chi_2^{\prime}}(t\p)$ at time $t\p$ to $p^{\chi_1}_{\chi_2}(t)$ at time $t$. It is given by the sum over all irreducible blocks on the Keldysh contour which correspond to the described transition.
The elements of $\mde$ are differences of the eigenenergies defined as $\left(\mde(t)\right)^{\chi_1\chi_1\p}_{\chi_2\chi_2\p}=\left(E_{\chi_{1}}(t)-E_{\chi_2}(t)\right)\delta_{\chi_1\chi_1\p}\delta_{\chi_2\chi_2\p}$. The second term $ \left[\bm{T}^\dagger\dot{\bm{T}},\bm{p}(t)\right]$ originates from the time dependence of the transformation $\bm{T}$, and $\dot{\bm{T}}$ denotes the time-derivative of $\bm{T}$.    
In the following adiabatic expansion and the expansion in the tunnel-coupling strength, we follow the lines of Ref. \onlinecite{splettstoesser_adiabatic_2006}.

\subsection{Adiabatic expansion\label{adiabatic}}
In the limit of slow variation of the system parameters, such that the duration of one pumping cycle, $\mathcal{T}=2\pi/\Omega$, is much larger than the dwell time of an electron in the quantum dot, we can perform an adiabatic expansion of Eq.~\eqref{master}, which is equivalent to an expansion of all time dependencies around the final time $t$ and to systematically keep all contributions that contain one time derivative.
For this, we first do a Taylor expansion of the reduced density matrix around the finite time $t$, i.e., $\bm{p}(t^\prime)\rightarrow \bm{p}(t)+(t^\prime-t)\ddt \bm{p}(t)$.
We then expand the kernel and the density matrix in the pumping frequency, i.e., 
$\bm{p}(t)\rightarrow \bm{p}^{(i)}_t+\bm{p}^{(a)}_t$ and $\K(t,t^\prime)\rightarrow \K^{(i)}_t(t-t^\prime)+\K^{(a)}_t(t-t^\prime)$.
The instantaneous order, indicated by the index $(i)$, describes the limit where all system parameters are frozen at time $t$. 
The adiabatic correction, labeled by $(a)$, contains one time derivative, i.e., it collects all contributions to first order in the pumping frequency $\Omega$. 
The difference in the eigenenergies of the isolated dot, $\mde(t)$, is of instantaneous order, while $\bm{T}^\dagger\dot{\bm{T}}(t)$ belongs to the adiabatic correction.

Since both $\K^{(i)}$ and $\K^{(a)}$ depend only on the difference $t-t^\prime$, it is convenient to perform the Laplace transform $F(z)=\int^{t}_{-\infty}\dtp e^{-z(t-t^\prime)}F(t-t^\prime)$.
Using the short notations $\K^{(i/a)}_t=\K^{(i/a)}_t(z=0^+)$ and $\partial\K^{(i)}_t=(\partial \K^{(i)}_t(z)/\partial z)|_{z=0^+}$, the kinetic equation reads
\begin{align}
 0&=\left(\K^{(i)}_t-\frac{i}{\hbar}\mde\right) \bm{p}^{(i)}_t \label{MEqi} \, ,
\end{align}
in instantaneous order, and
\begin{align}
 \ddt \bm{p}^{(i)}_t=& \left( \K^{(i)}_t -\frac{i}{\hbar}\mde\ \right) \bm{p}^{(a)}_t- \left[\bm{T}^\dagger\dot{\bm{T}}\, ,\bm{p}^{(i)}_t\right]\nonumber\\&+\K^{(a)}_t \bm{p}^{(i)}_t+\partial\K^{(i)}_t\ddt \bm{p}^{(i)}_t \label{MEqa}
\end{align}
for the adiabatic correction.
The normalization condition for the density matrix is expressed as ${\rm Tr} \, \bm{p}^{(i)}_t =1$ and ${\rm Tr} \, \bm{p}^{(a)}_t=0$.

\subsection{Expansion in the tunnel-coupling strength\label{tunnel}}

In addition to the adiabatic expansion, we perform a perturbation expansion in the tunnel-coupling strength $\Gamma$. 
For a systematic expansion of the kinetic equations, we need to analyze the term $\mde\ \bm{p}^{(i/a)}$.
It vanishes for all diagonal matrix elements of $\bm{p}^{(i/a)}$.
The off-diagonal matrix elements, associated with coherent superpositions, are only nonzero when the superposition is not forbidden by conserved quantum numbers and when the energy difference of the corresponding states is smaller or of the order of $\Gamma$.
Therefore, we count all contributing matrix elements of $\mde$ to be of the order of $\Gamma$.

The expansion of the kernels, $\K_t^{(i/a)}=\sum_{n=1}^{\infty}\K_t^{(i/a,n)}$, starts to first order in $\Gamma$ and the instantaneous order of the reduced density matrix to zeroth order, $\bm{p}_t^{(i)}=\sum_{n=0}^{\infty}\bm{p}_t^{(i,n)}$. To properly match the powers of $\Gamma$ in Eq.~\eqref{MEqa}, the adiabatic correction of the reduced density matrix, $\bm{p}_t^{(a)}=\sum_{n=-1}^{\infty}\bm{p}_t^{(a,n)}$, has to start to minus first order.\cite{splettstoesser_adiabatic_2006}

In the following, we consider the limit of weak tunnel coupling, $\Gamma \ll k_BT$, for which we restrict ourselves to the lowest-order contributions in $\Gamma$.
The instantaneous part of the kinetic equation starts to first order in $\Gamma$, 
\begin{align} 
	0&=\left(\K^{(i,1)}_t-\frac{i}{\hbar}\mde\right) \bm{p}_t^{(i,0)}\, , \label{eq:masteri}
\end{align}
with normalization ${\rm Tr}\,  \bm{p}_t^{(i,0)}=1$.
For the adiabatic correction, the expansion of Eq.~(\ref{MEqa}) to lowest (zeroth) order in $\Gamma$ yields
\begin{align} 
	\ddt \bm{p}^{(i,0)}_t =\left(\K^{(i,1)}_t-\frac{i}{\hbar}\mde\right) \bm{p}_t^{(a,-1)}\, ,\label{eq:mastera} 
\end{align}
with ${\rm Tr}\,  \bm{p}_t^{(a,0)}=0$.
All other terms appearing on the right-hand side of Eq.~(\ref{MEqa}) are of higher order in $\Gamma$ and drop out.
This is immediately obvious for the last two terms in Eq.~(\ref{MEqa}).
But also $\left[\bm{T}^\dagger\dot{\bm{T}},\bm{p}^{(i,0)}_t\right]$ drops out in the absence of any bias voltage. In this case, $\bm{p}^{(i,0)}_t$ is given by the equilibrium distribution, which is diagonal with matrix elements being determined by Boltzmann factors. Since energy differences, $\mde$, of states for which coherent superpositions are allowed are of the order of $\Gamma$, the difference of the corresponding occupation probabilities for these states is also of the order of $\Gamma$ and, therefore, vanishes in the perturbation expansion.
This means that the matrix elements \begin{align}\left(\left[\bm{T}^\dagger\dot{\bm{T}},\bm{p}_t^{(i,0)}\right]\right)^{\chi_1}_{\chi_2}=\left(p^{(i,0)}_{t\ \chi_2}-p^{(i,0)}_{t\ \chi_1}\right)\left(\bm{T}^\dagger\dot{\bm{T}}\right)^{\chi_1}_{\chi_2}\end{align} vanish for all combinations of $\chi_1$ and $\chi_2$ which are needed in the kinetic equation.

\subsection{Pumped charge and pumped spin\label{pumped}}
The pumped current and the pumped spin current from the dot into the left lead are given by
\begin{align}
I_\text{L}(t)&= e \int_{-\infty}^t\dtp \trace{\K^{\text{L}}_Q(t,t\p)\:\bm{p}(t\p)},\\
S_\text{L}(t)&= \frac{\hbar}{2} \int_{-\infty}^t\dtp \trace{\K^{\text{L}}_S(t,t\p)\:\bm{p}(t\p)},
\end{align} respectively.
Here, we have introduced $\K^{\text{L}}_{Q/S}(t,t\p)=\sum_{q} (q_\up\pm q_\down) \K^{\text{L} q_\up q_\down}(t,t')$ where $\K^{\text{L} q_\up q_\down}(t,t')$ only contains those diagrams of $\K(t,t')$ in which $q_\s$ electrons with spin $\s$ enter the left lead, i.e., in which the number of lines for lead $L$ and spin $\s$ going from the upper to the lower contour minus the number of lines from the lower to the upper contour is $q_\s$. 

Analogously to the expansion of the kinetic equation, we perform the adiabatic expansion and the perturbation expansion in the tunnel-coupling strength for the pumped charge and spin current. 
To lowest order we get
\begin{align}
I^{(a,0)}_\text{L}(t)&= e \trace{\left(\K^{\text{L}}_{Q,t}\right)^{(i,1)}\:\bm{p}_t^{(a,-1)}}, \label{eq:current1}\\
S^{(a,0)}_\text{L}(t)&= \frac{\hbar}{2} \trace{\left(\K^{\text{L}}_{S,t}\right)^{(i,1)}\:\bm{p}_t^{(a,-1)}}.\label{eq:current2}
\end{align}
The pumped charge and the pumped spin per pumping cycle is obtained by integration, $Q= \int^\mathcal{T}_0 \dt I^{(a,0)}_\text{L}(t)$ and $\Sigma=\int^\mathcal{T}_0 \dt S^{(a,0)}_\text{L}(t)$. 
The diagrammatic rules to calculate analytically $\K^{(i,1)}$ can be found in Appendix~\ref{rules}. \cite{schoeller_mesoscopic_1994,knig_resonant_1996,knig_zero-bias_1996,splettstoesser_adiabatic_2006,sothmann_transport_2010} 
After having determined $\K^{(i,1)}$, we obtain the adiabatic correction to the reduced density matrix, $\bm{p}_t^{(a,-1)}$, by solving the kinetic equations \eqref{eq:masteri} and \eqref{eq:mastera}. 
Those, then, enter Eqs.~\eqref{eq:current1} and \eqref{eq:current2} for the pumped charge and spin currents.

\subsection{Weak pumping\label{weak}}
We split the energy of the orbital levels into the time-averaged part, $\bar{\e}_\n=\frac{1}{\mathcal{T}}\int_0^{\mathcal{T}} \text{d}t\; \e_\n(t)$, and the deviation $\delta\e_\n(t)$:
\begin{align}
\e_1(t)&=\bar{\e}_1+\delta\e_1(t), \\
\e_2(t)&=\bar{\e}_2+ \delta\e_2(t).
\end{align}
In the limit of weak pumping, the time-dependent part of the pumping parameters is small compared to other energy scales of the system such as tunnel-coupling strength and temperature, $\delta\epsilon_\n(t)\ll \Gamma,k_BT$. Hence, we can expand the pumped charge, $Q$, and pumped spin, $\Sigma$, to lowest (bilinear) order in $\e_1(t)$ and $\e_2(t)$.
For adiabatic pumping, the pumped charge (spin) is proportional to the area enclosed by the path of $(\e_1(t),\e_2(t))$ in the parameter space during one pumping cycle.
Therefore, a phase difference is necessary to gain finite pumped charge (spin).
The enclosed area is given by $\eta=\int_0^{\mathcal{T}}\text{d}t\; \delta \e_1(t)\; \partial_t \delta \e_2(t)$. 
All results in Sec.~\ref{res} are calculated in the weak-pumping limit.

\section{Isospin transformation\label{isospin}}

For each matrix element $p^{\chi_1}_{\chi_2}$ that needs to be considered (all diagonal ones and those off-diagonal ones that describe possible coherent superpositions), there is one kinetic equation. 
It is often convenient to transform the reduced density matrix such that only linear combinations of the $p^{\chi_1}_{\chi_2}$'s appear, which allows for a straightforward physical interpretation.
In the context of spin transport through a single-level quantum dot with ferromagnetic leads, it is advantageous to formulate the kinetic equations separately for the occupation probability of zero, one or two electrons on the quantum dot, and the three components of the spin on the dot.~\cite{koenig_interaction-driven_2003,braun_theory_2004,splettstoesser_adiabatic_2008}
The vector character of the spin accounts for both a spin imbalance along a given axis and the coherent dynamics of the accumulated spin.
One virtue of such a transformation lies in the fact that it is possible to write the kinetic equation in a (spin-)coordinate-free form, which does not depend on the choice of the spin quantization axis.

A similar transformation can also be used for the orbital degree of freedom in systems in which coherent superpositions of the occupation of different orbitals appear.
These superpositions are conveniently described by defining an isospin.
This has been done before for several double-dot systems.~\cite{wunsch_probing_2005,legel_generation_2007,urban_tunable_2009,riwar_charge_2010}
We will introduce such an isospin description now for the system under consideration. 

In this paper, we focus on the limits of $U=0$ and $U=\infty$. 
In the first case, $U=0$, the Hilbert space is $16$-dimensional, i.e., the reduced density matrix is a $16 \times 16$ matrix.
However, since we choose the spin-quantization axes along the direction of the \so~field, the Hamiltonian divides into two independent spin channels.
As a result, the reduced density matrix can be written as a direct product of the $4\times 4$ density matrices for spin up and spin down, $ \bm{p}_{U=0} = \left(\bm{p}_{U=0}\right)_\up  \otimes \left(\bm{p}_{U=0}\right)_\down $.
In the basis $\{ \ket{0}, \ket{1}, \ket{2}, \ket{d} \}_\s$ that corresponds, for each spin $\s$, to the occupation of none of the orbitals, of orbital 1, of orbital 2, and of both orbitals, respectively, the reduced density matrix reads
\begin{align}
\left(\bm{p}_{U=0}\right)_\s =\begin{pmatrix}p_0 & 0& 0 &0\\ 0 & p_{1}& p_{2}^{1} &0\\ 0 & p_{1}^{2}& p_{2} &0\\ 0 & 0& 0 &p_d \end{pmatrix}_\s\, .
\end{align}
Note that, in order to keep the notation simple, we put the index $\s$ only once at the matrix indicating the $\s$-dependence of each of the matrix elements.

For $U=\infty$ the dot is either singly occupied or empty, i.e., the Hilbert space is five-dimensional.
The reduced density matrix takes the form
\begin{align}
\bm{p}_{U=\infty}=\begin{pmatrix}p_0 & 0& 0 &0&0 \\ 0 & p_{1\up}& p_{2\up}^{1\up}&0&0 \\ 0 & p_{1\up}^{2\up}& p_{2\up}&0&0\\ 0 &0&0& p_{1\down}& p_{2\down}^{1\down} \\ 0&0&0 & p_{1\down}^{2\down}& p_{2\down} \end{pmatrix}\, .
\end{align}
Note that, here, $p_0$ is the probability that the dot is not occupied with either spin, while for $U=0$ we used $(p_0)_\s$ for the probability that the dot is not occupied with spin $\s$, irrespective of the occupation of spin $-\s$.

To describe the coherent superposition associated with the off-diagonal matrix elements, it is convenient to introduce, for each physical spin, an isospin operator $ \bm{\hat I}_\s $ with quantum-statistical expectation value $\bm{I}_\s = \langle \bm{\hat I}_\s \rangle$.
Choosing the coordinate system for the isospin such that $\ket{1\s}$ and $\ket{2\s}$ are the eigenstates of $\hat I_z^\s$, we get $I_x^\s=\left(p^{1\s}_{2\s}+p^{2\s}_{1\s}\right)/2$, $I_y^\s=i\left(p^{1\s}_{2\s}-p^{2\s}_{1\s}\right)/2$, and $I_z^\s=\left(p_{1\s}-p_{2\s}\right)/2$. 
Since ultimately we aim at a coordinate-free form of the kinetic equations, we abbreviate the $z$-axis chosen above by the normalized vector $\bm{n}$, i.e., $\ket{1\s}$ and $\ket{2\s}$ are the eigenstates of $\bm{\hat I}_\s\cdot \bm{n}$.

The isospin direction $\bm{n}$ characterizes the eigenstates of the isolated quantum dot in the absence of \so~coupling.
The \so~coupling, however, couples the two orbitals.
As a consequence, the dot eigenstates 
  \begin{align}
 \begin{pmatrix}\ket{+\s}\\ \ket{-\s}\end{pmatrix}= \bm{T}_\s \begin{pmatrix}\ket{1\s} \\ \ket{2\s} \end{pmatrix}
 \end{align}
for single occupation with spin $\s$ are linear combinations of the two orbitals $\ket{1\s}$ and $\ket{2\s}$, given by the transformation
\begin{align}
\bm{T}_\s= \frac{1}{\sqrt{2\xi(\xi+\dep)}}
\begin{pmatrix}
  \xi+\dep & i \s \aso \\
  i \s \aso & \xi+\dep 
\end{pmatrix}\, .
\end{align}
The corresponding eigenenergies are $E_{\pm} = \e \pm \xi$, with the mean dot level $\e=\left(\e_1+\e_2\right)/2$ and $\xi=\sqrt{\dep^2+\aso^2}$.
The transformation depends on the spin $\s$, the level spacing of the both orbitals, $\dep=(\e_1-\e_2)/2$, and the strength of the \so~coupling, $\aso$.
We consider the regime where $\aso$ and $\dep$ are of order $\Gamma$. Therefore, the level spacing $2\xi$ of the eigenenergies $E_\pm$ is also of order $\Gamma$.
As we pump on both energies, $\e_1(t)$ and $\e_2(t)$, the transformation $\bm{T}_\s(t)$ and the eigenenergies $E_{\pm}(t)$ are time dependent.

\begin{figure}
	\includegraphics[width=.45\textwidth]{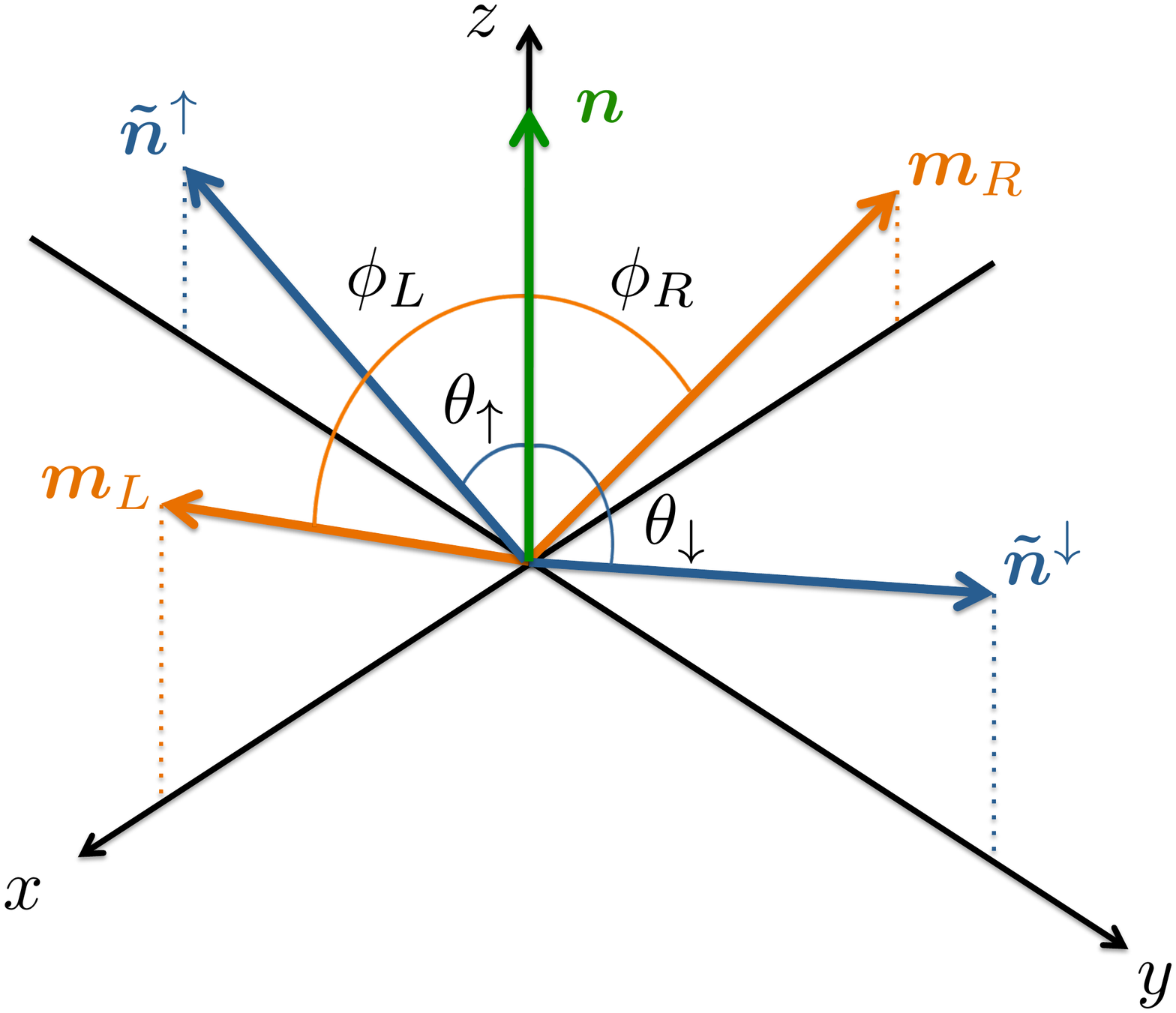}
	\caption{\label{fig:mn}(Color online) Scheme of different relevant isospin quantization axes. The vector $\bm{n}$ represents the quantization where the orbital levels $\ket{1\s}$ and $\ket{2\s}$ are the eigenstates of the $\hat{I}_z$ operator of the isospin. The two axes $\tilde{\bm{n}}_\s$ are the quantization axes where the eigenstates of $\hat{I}_z$ are the eigenstates of $H^{\text{dot}}$ for single occupation. In a ferromagnetic analogy, the leads are fully isospin polarized along the axes $\bm{m}_\la$.}
\end{figure}

The unitary transformation $\bm{T}_\s$ corresponds to a rotation about the $x$-axis with the spin-dependent angle 
\begin{align}
	\theta_\s=- \s \arcsin \left(\frac{\aso}{\xi}\right)
\end{align} 
in isospin space.
This means that the dot eigenstates $\ket{\pm\s}$ are eigenstates to the isospin projection $\bm{\hat I}_\s\cdot \bm{\tilde{n}}_{\s}$ along the direction $\bm{\tilde{n}}_{\s}$ that is obtained from $\bm{n}$ by the above mentioned rotation (see Fig.~\ref{fig:mn}).

The tunneling Hamiltonian couples the lead-electron states to both orbitals, i.e., to a linear combination of $\ket{1\s}$ and $\ket{2\s}$.
To diagonalize the tunneling from and to lead $\la$, we employ the unitary transformation 
\begin{align}
 \bm{F}_\la&= \frac{1}{\sqrt{V_{\la1}^2+V_{\la2}^2}}\begin{pmatrix}V_{\la1}&V_{\la2}\\
-V_{\la2}&V_{\la1}\end{pmatrix}\, .
\end{align}
In isospin space, this transformation corresponds to a rotation about the $y$-axis with angle
\begin{align}
	\phi_\la=- \arcsin \left( \frac{2V_{\la1}V_{\la2}}{V_{\la1}^2+V_{\la2}^2} \right) \, . \label{eq:Phila}
\end{align} 
Applying this rotation on $\bm{n}$ generates the direction $\bm{m}_\la$ (see Fig.~\ref{fig:mn}) which has the following physical interpretation: Only dot electrons with $\ket{+}_{\bm{\hat I}_\s\cdot \bm{m}_{\la}}$ isospin projection along $\bm{\hat I}_\s\cdot \bm{m}_{\la}$ couple to reservoir $\la$, while the $\ket{-}_{\bm{\hat I}_\s\cdot \bm{m}_{\la}}$ isospin projection is decoupled from the lead.\cite{legel_generation_2007} 
Therefore, in a ferromagnetic analogy, the leads are full isospin polarized with polarization along  $\bm{m}_\la$.

First, we write the kinetic equations \eqref{eq:masteri} and \eqref{eq:mastera} in the basis $\{ \ket{0}, \ket{+}, \ket{-}, \ket{d} \}_\s$ for the $U=0$ limit and  $\{ \ket{0}, \ket{+\up}, \ket{-\up}, \ket{+\down}, \ket{-\down} \}$ for $U=\infty$. Those kinetic equations are treated perturbatively to first order in the tunnel-coupling strength $\Gamma$. As described above, we count both $\aso$ and $\dep$ as one order in $\Gamma$. The elements of the kernel $\K^{(i,1)}$ are calculated by the rules in Appendix~\ref{rules}. Including the isospin in the formulation of the kinetic equation, the system is fully described by the occupation probabilities of the dot and the expectation values of the isospins. In particular, we perform the transformation from $\{p_0,p_{+},p_{-},p^{-}_{+},p^{+}_{-},p_d\}_\s$ to $\{p_0,p_s,p_d,\bm{I}\}_\s$ in the limit of vanishing \ci. The probabilities describing the occupation of the dot with spin $\s$ are $\left(p_0,p_s,p_d\right)_\s$ for empty, $p_0$, single, $p_s=p_1+p_2$, and double occupation, $p_d$. In the limit of strong \ci, $U=\infty$, the transformation reads $\{p_0,p_{+\up},p_{-\up},p^{-\up}_{+\up},p^{+\up}_{-\up},p_{+\down},p_{-\down},p^{-\down}_{+\down},p^{+\down}_{-\down}\}$ to  $\{p_0,p_\up,p_\down,\bm{I}_\up,\bm{I}_\down\}$. The relevant occupation probabilities are $\left(p_0,p_\up,p_\down\right)$ with $p_\s=p_{1\s}+p_{2\s}$ being the possibility that the dot is occupied by a single electron with spin $\s$.  We identify in the resulting kinetic equations the vectors $\bm{m}_\la$ and $\bm{\tilde{n}}_\s$ and get, thus, a representation that is independent of the choice of basis. In the limit of $U=0$, we get
\begin{subequations}
\label{eq:Dia:MEQU0}
\begin{align}
	\ddt \begin{pmatrix}p_0\\p_s\\p_d\end{pmatrix}_\s &=\frac{\Gamma}{\hbar} \begin{pmatrix}- f  & \frac{1-f}{2}&0\\  f & -\frac{1}{2}&1-f \\ 0 & \frac{f }{2}&-(1-f)  \end{pmatrix} \begin{pmatrix}p_0\\p_s\\p_d\end{pmatrix}_\s\nonumber\\ &+ \frac{\Gamma}{\hbar}   \begin{pmatrix}1-f \\2f-1 \\-f \end{pmatrix} \left( \bm{I}_\s \cdot \bm{\bar{m}} \right)\, ,  \label{eq:Dia:MEQU01}
\\ 
\ddt \bm{I}_\s &=  \frac{\Gamma}{\hbar}  \left(\frac{f}{2}p_0+\frac{2f-1}{4}p_s-\frac{1-f}{2}p_d\right)_\s \bm{\bar{m}} \nonumber \\&-\frac{\Gamma}{2\hbar} \bm{I}_\s + \bm{I}_\s \times  \bm{B}_{\s}
\label{eq:Dia:MEQU02}\, ,
\end{align}
\end{subequations}
where $f=f(\e)$ is the Fermi function at energy $\e$. As the difference of the eigenenergies, $2\xi$,  is of order $\Gamma$, we have to drop $2\xi$ in terms which are already linear in $\Gamma$. Therefore, the Fermi function, $f$, depends here only on the mean level position, $\e$, since every term which includes the Fermi function is linear in $\Gamma$.  
In the equations for the probabilities, the isospin projections along the directions defined by the leads enter in the weighted average
\begin{align}
	\bm{\bar{m}}=\frac{\Gamma_\text{L}}{\Gamma}\bm{m}_\text{L}+\frac{\Gamma_\text{R}}{\Gamma}\bm{m}_\text{R} \, ,
\end{align}
with $\Gamma_\la=\sum_\n \Gamma_{\la\n\n}$.
The isospin projection direction given by the \so~coupling, on the other hand, gives rise to a precession term about the effective field
\begin{align}
 \bm{B}_{\s}&=-\frac{ 2 \xi}{\hbar}\bm{\tilde{n}}_{\s}
\end{align}
in the equation for the isospin. 
This effective field is the only place where the \so~coupling enters the kinetic equations.
Equations~\eqref{eq:Dia:MEQU01} and \eqref{eq:Dia:MEQU02} represent both the instantaneous order and the adiabatic correction of the kinetic equation.
For the first case, one needs to set the left-hand side to zero and add the index $(i,0)$ to the isospin and the occupation probabilities on the right-hand side.  
(Note that the instantaneous part of the isospin vanishes in lowest order in $\Gamma$, $\bm{I}_\s^{(i,0)}=0$.)
For the the second case, we need to add the index $(i,0)$ on the left-hand side and $(a,-1)$ on the right-hand side,
respectively.

In the limit of strong \ci, $U=\infty$, the kinetic equations read
\begin{subequations}
\label{eq:Dia:MEQUI}
\begin{align}
\ddt \begin{pmatrix}p_0\\p_\up\\p_\down\end{pmatrix}&=\frac{\Gamma}{\hbar}  \begin{pmatrix}-2 f & (1-f)/2&(1-f)/2\\  f & -(1-f)/2&0\\ f & 0&-(1-f)/2 \end{pmatrix} \begin{pmatrix}p_0\\p_\up\\p_\down\end{pmatrix}\nonumber \\&+\frac{\Gamma}{\hbar} (1-f)  \begin{pmatrix}\left( \bm{I}_{\up} \cdot \bm{\bar{m}} \right)+\left( \bm{I}_{\down} \cdot \bm{\bar{m}} \right)\\-\left( \bm{I}_{\up} \cdot \bm{\bar{m}} \right)\\-\left( \bm{I}_{\down} \cdot \bm{\bar{m}} \right)\end{pmatrix}\, ,\label{eq:Dia:MEQUI1} \\
\ddt \bm{I}_{\s}&= \frac{\Gamma}{\hbar}  \left(\frac{f}{2}p_0-\frac{1-f}{4}p_\s\right) \bm{\bar{m}} \nonumber \\&-\frac{\Gamma}{\hbar}  \frac{1-f}{2} \bm{I}_{\s}
 +  \bm{I}_{\s} \times \left( \bm{B}_{\s} + \bm{B}_{U}\right)
 \label{eq:Dia:MEQUI2}\, .
\end{align}
\end{subequations}

In addition to the effective field $\bm{B}_\s$ generated by the \so~coupling, we identified here another effective field $\bm{B}_U$ acting on the isospin. 
The latter appears as a consequence of the interplay between tunneling and Coulomb interaction. 
It is formally identical to the exchange field acting on the physical spin in quantum dots attached to ferromagnetic leads.\cite{braun_theory_2004,koenig_interaction-driven_2003,splettstoesser_adiabatic_2008}
In the limit of $U= \infty$, it is given by
\begin{align}
 \bm{B}_{U}&= \frac{\Gamma}{2\pi\hbar}    \bm{\bar{m}} \,\, {\rm Re} \int \dw \frac{f(\w)}{\e-\w+i0^+} 
 \nonumber \\
 	&= \frac{\Gamma}{2\pi\hbar} \left[ \ln \frac{\beta U_{\rm cutoff}}{2\pi} - {\rm Re} \, \Psi \left(\frac{1}{2}+i\frac{\beta \e}{2\pi} \right) \right]
  \, ,\label{eq:Dia:BUFeld}
\end{align}
where $\Psi$ is the digamma function and we used $\beta = 1/k_BT$. The high-energy cutoff $U_{\rm cutoff}$ appearing in the second line guarantees convergence of the energy integral.\cite{schoeller_mesoscopic_1994,knig_resonant_1996,knig_zero-bias_1996} 
Physically it is provided by the smaller of the band width of the leads and the charging energy.

For practical calculations it is not necessary to use the basis-independent form of this isospin representation. It allows for a better physical understanding of the systems dynamics but for evaluating the pumped charge and spin as described in Sec.~\ref{method}, it is convenient to use the basis $\{ \ket{0}, \ket{1}, \ket{2}, \ket{d} \}_\s$ in the $U=0$ limit and  $\{ \ket{0}, \ket{+\up}, \ket{-\up}, \ket{+\down}, \ket{-\down} \}$ for $U=\infty$.  
 
\section{Results\label{res}}

In this section, we present the results for the adiabatically pumped charge (spin) in the weak-pumping regime. To calculate those, we use the formalism that has been introduced in Sec. \ref{method}. We integrate the pumped charge and spin currents, Eqs.~\eqref{eq:current1} and \eqref{eq:current2}, over one pumping cycle and obtain the pumped charge and pumped spin per pumping cycle. In order to simplify the time dependence of the pumped currents, we make use of the weak pumping limit (see Sec.~\ref{weak}) and expand the integrand up to bilinear order in the pumping parameters, $\e_1(t)$ and $\e_2(t)$. In this case, all results are proportional to the area, $\eta$, enclosed in the pumping-parameter space. We normalize our results by $\eta$ and, thus, they are independent of the exact path in parameter space. 

To analyze the effect of \ci, we compare results for noninteracting electrons, $U=0$, with the limit of strong \ci. The latter is realized by setting $U=\infty$ in the Hamiltonian and, thereby, suppressing occupation of the quantum dot with more than one electron. Furthermore, finite \ci~influences the amplitude of the exchange field, $\bm{B}_U$, via the high-energy cutoff, $U_{\rm cutoff}$. In all calculations, we set $U_{\rm cutoff}=100 k_BT$.
We assume weak tunnel coupling between quantum dot and leads, $\Gamma \ll k_BT$, i.e., we restrict the calculation to lowest order in the tunnel-coupling strength $\Gamma$.
If not stated otherwise, the \so-coupling strength is $\aso=\Gamma/10$.

For $U=0$, the results of this paper can be compared with calculations that include higher orders in $\Gamma$. For example, Brosco\etal~have studied the two-level quantum dot with \so~coupling and vanishing \ci~in the limit of zero temperature.\cite{brosco_prediction_2010}
 Calculations to all orders in $\Gamma$ can be done, e.g., with a scattering matrix approach,\cite{bttiker_current_1994,brouwer_scattering_1998,avron_geometry_2000} which is equivalent to an approach that is based on a formula relating the pumped current to the instantaneous dot Greens functions.\cite{splettstoesser_adiabatic_2005} The latter is, in general, extendable to finite interaction.  

In this section, we study the dependence of the pumped charge (spin) on various parameters: the strength of the \so~coupling, $\aso$, the tunnel coupling to the leads, $V_{\la\n}$, and the time-averaged dot levels, $\bar{\e}_\n$. It is convenient to parametrize the latter by the time-averaged mean dot level, $\be=(\overline{\e}_1+\overline{\e}_2)/2$, and the averaged spacing of both orbital levels, $\bdep=(\overline{\e}_1-\overline{\e}_2)/2$.

In the regime under consideration, the temperature appears only in two ways. 
First, since the mean energy level only appears in the combination $\beta \be$, the temperature provides the energy scale on which variation of the mean level energy changes the pumped charge (spin).
Second, the absolute value of the pumped charge and pumped spin are proportional to $\left(k_BT\right)^{-1}$.
Therefore, all plots are normalized accordingly.     
The dependences of the pumped charge (spin) on the other parameters are not affected by temperature.   

The tunnel coupling of the two dot orbitals to the left and the right lead is defined by four real tunnel-matrix elements, $V_{\la\n}$. If the tunnel-matrix elements are equal for the coupling to the left and right lead, $V_{\text{L}\n}=V_{\text{R}\n}$, then, for symmetry reasons, there will be no pumping transport via variation of the quantum dot's levels.  
To achieve pumping, the left-right symmetry needs to be broken by changing either the magnitude or the sign of one the tunnel couplings.
We find it convenient to parametrize the tunnel-matrix elements by angles $\phi_\la$, which have been introduced in the previous section (see Eq. \eqref{eq:Phila}).  The tunnel-matrix elements then are given by the relations $V_{\la1}=\sqrt{\frac{\Gamma_{\la}}{2\pi\rho}} \cos\frac{\phi_\la}{2}$ and $V_{\la2}=\sqrt{\frac{\Gamma_{\la}}{2\pi\rho}} \sin\frac{\phi_\la}{2}$. For $\phi_\la=\pi/2$ both orbital levels are coupled symmetrically to lead $\la$, i.e., $V_{\la1}=V_{\la2}$, and for $\phi_\la=-\pi/2$ the orbitals are coupled antisymmetrically, $V_{\la1}=-V_{\la2}$. 
The necessary condition to get a finite pumped charge (spin) is $\phi_\text{L} \neq \phi_\text{R}$, since $\phi_\text{L}=\phi_\text{R}$ (even for  $\Gamma_\text{L}\neq\Gamma_\text{R}$) leads automatically to an effective one parameter pumping without any finite pumped charge (spin) in the adiabatic limit.

\subsection{Charge and spin pumping}
\begin{figure*}
	\subfigure[\label{fig:QU0}~Pumped charge for $U=0$]{\includegraphics[width=.45\textwidth]{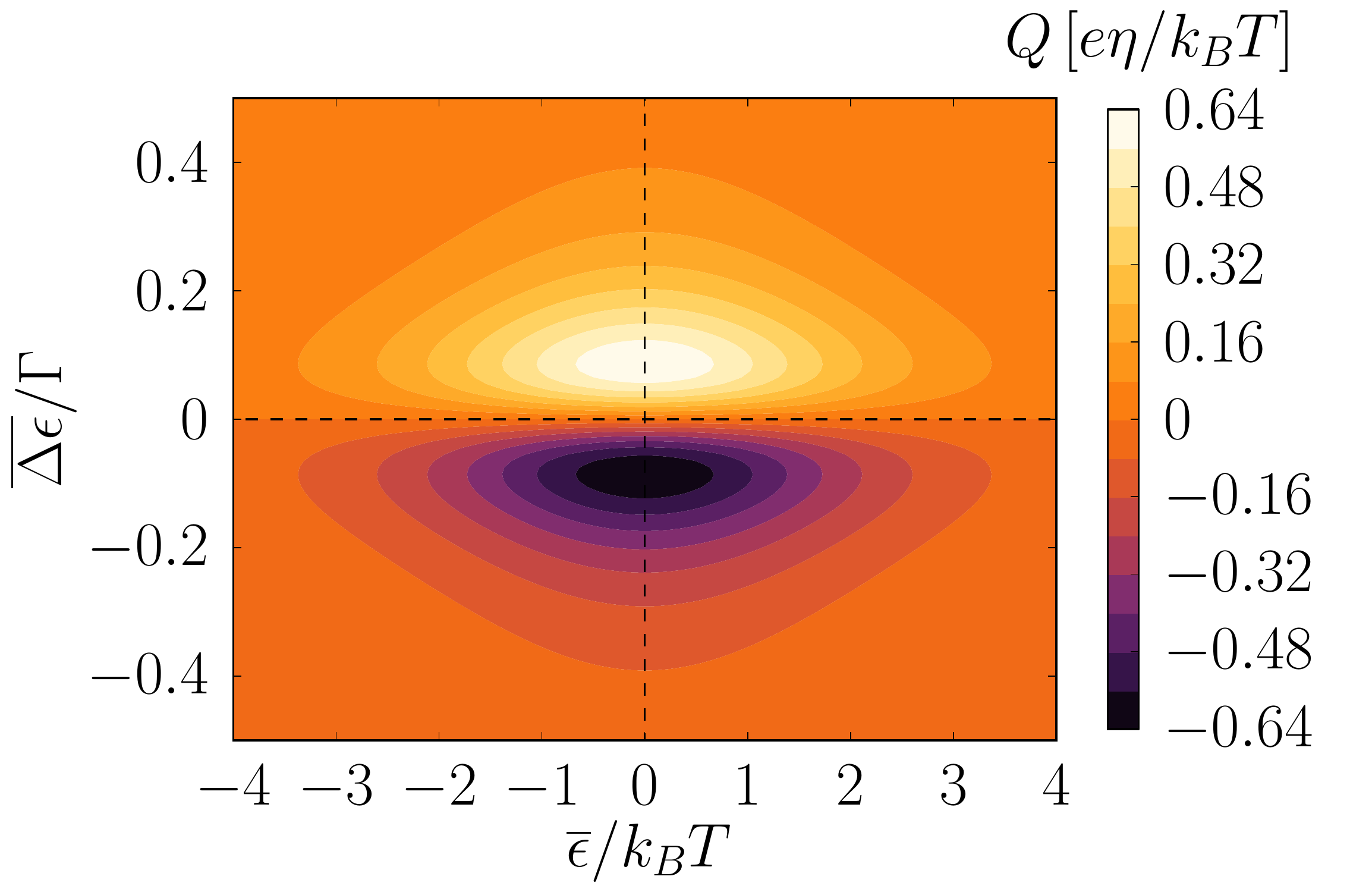}}
	\subfigure[\label{fig:QUI}~Pumped charge for $U=\infty$]{\includegraphics[width=.45\textwidth]{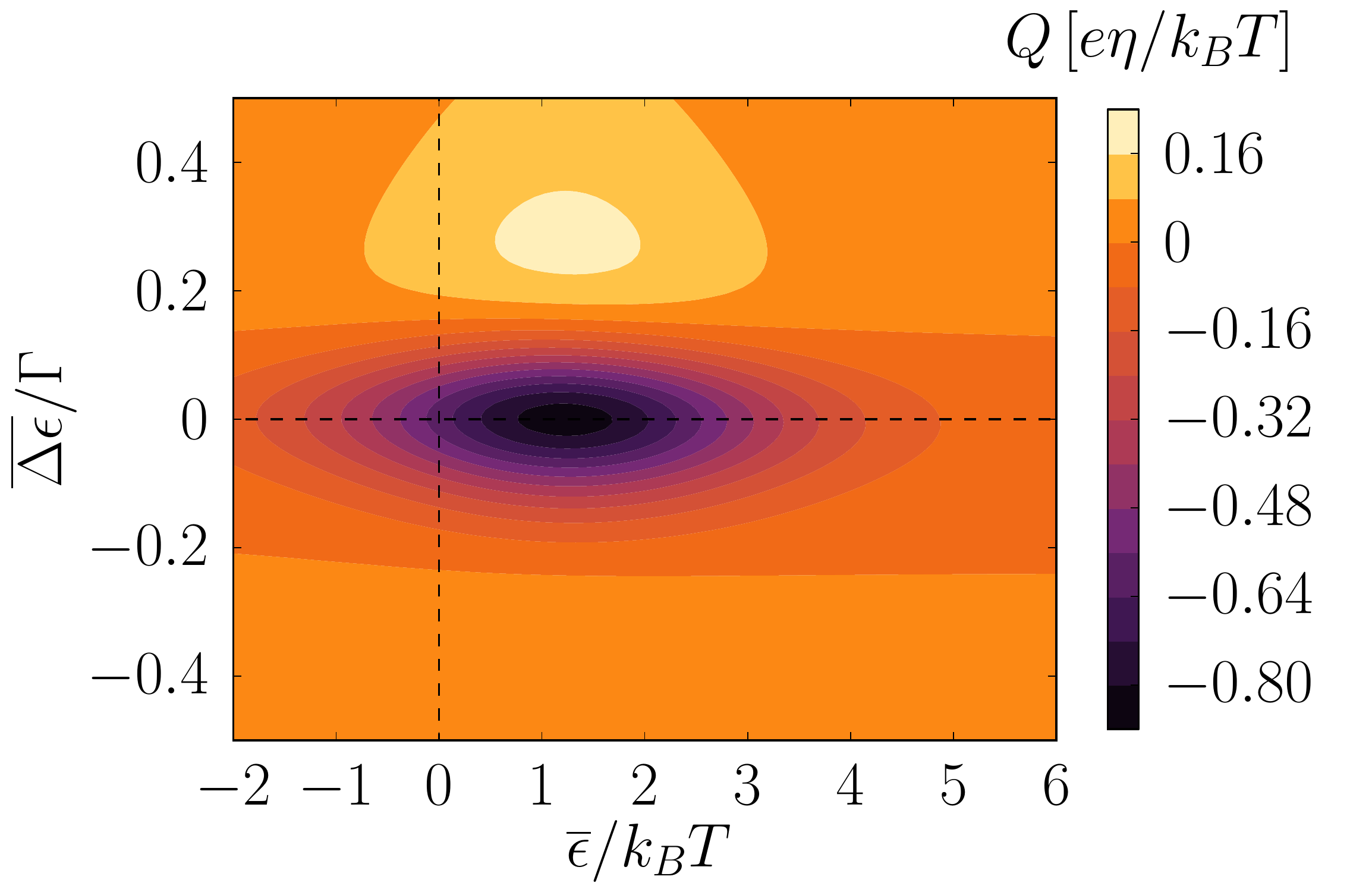}}
	\subfigure[\label{fig:SU0}~Pumped spin for $U=0$]{\includegraphics[width=.45\textwidth]{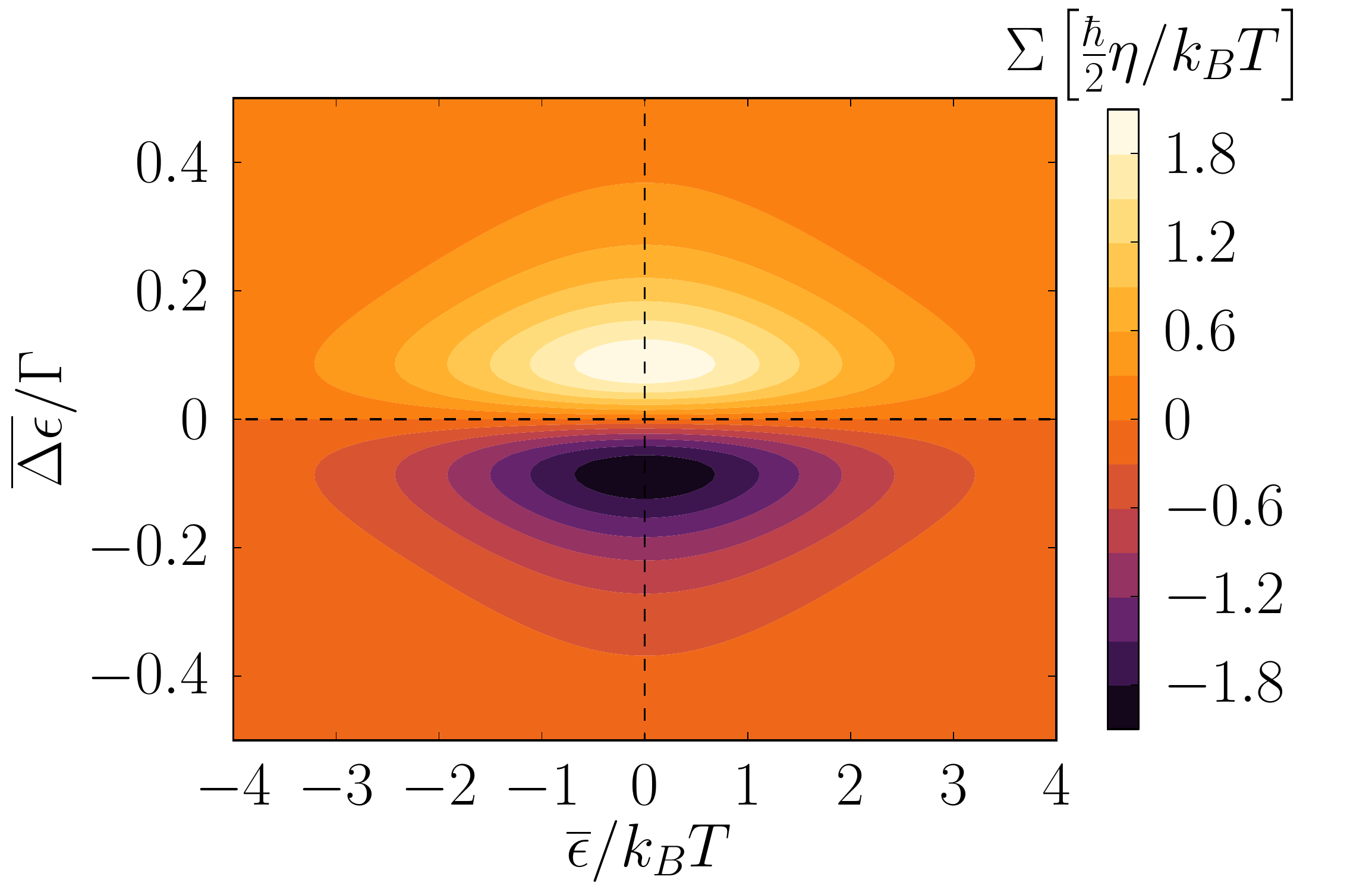}}
	\subfigure[\label{fig:SUI}~Pumped spin for $U=\infty$]{\includegraphics[width=.45\textwidth]{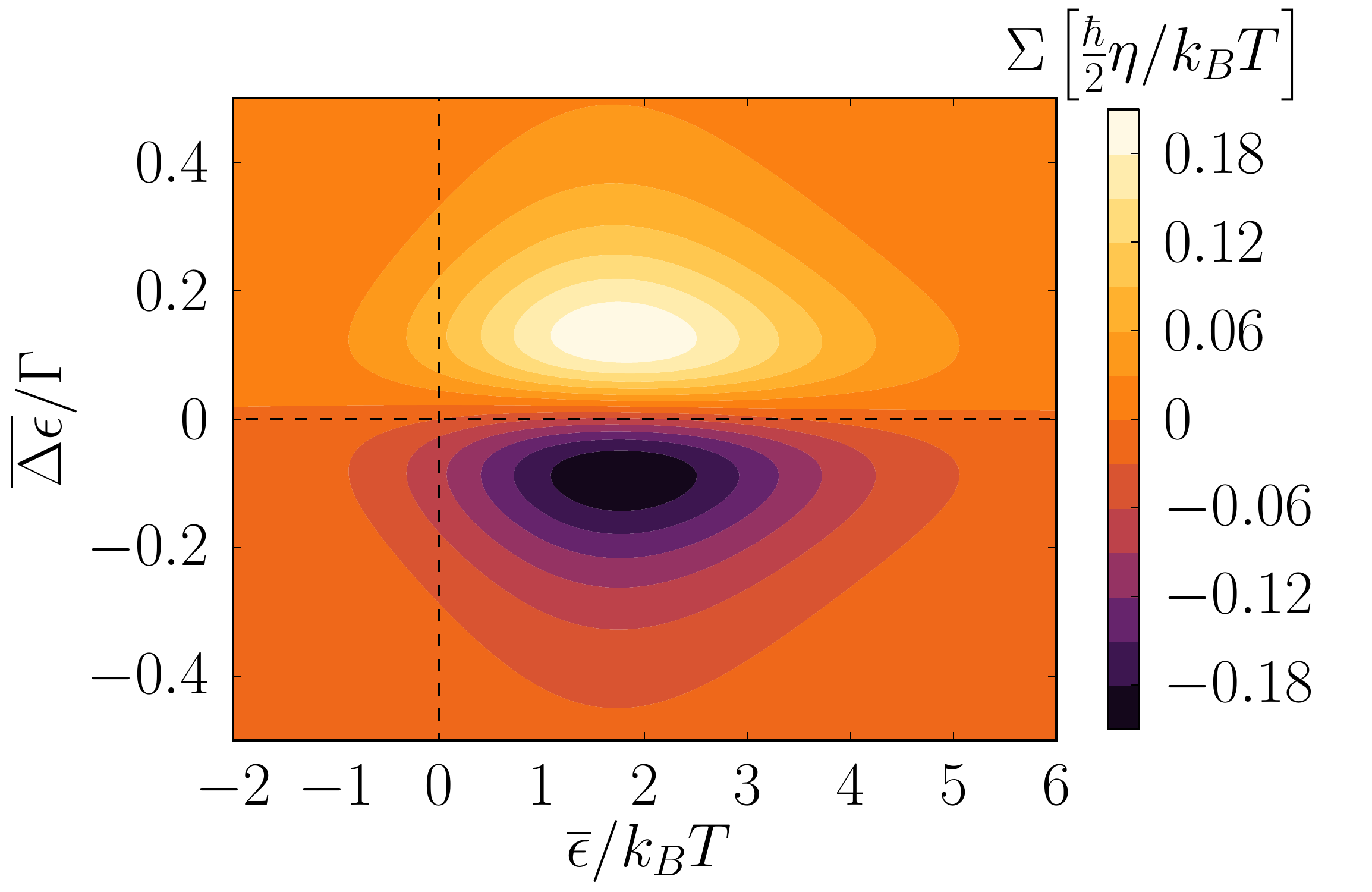}}
	\subfigure[\label{fig:PSU0}~Pure pumped spin for $U=0$]{\includegraphics[width=.45\textwidth]{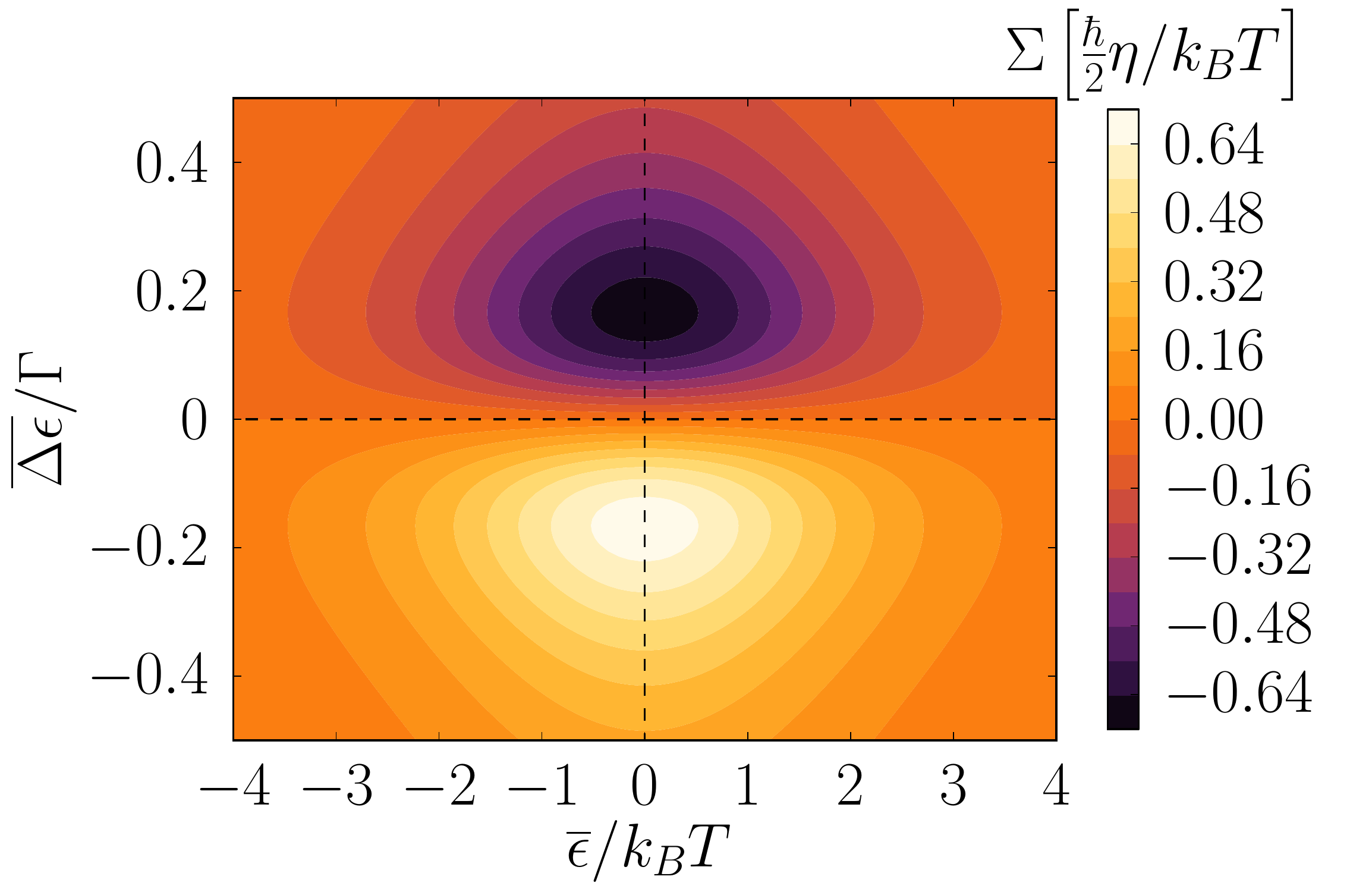}}
	\subfigure[\label{fig:PSUI}~Pure pumped spin for $U=\infty$]{\includegraphics[width=.45\textwidth]{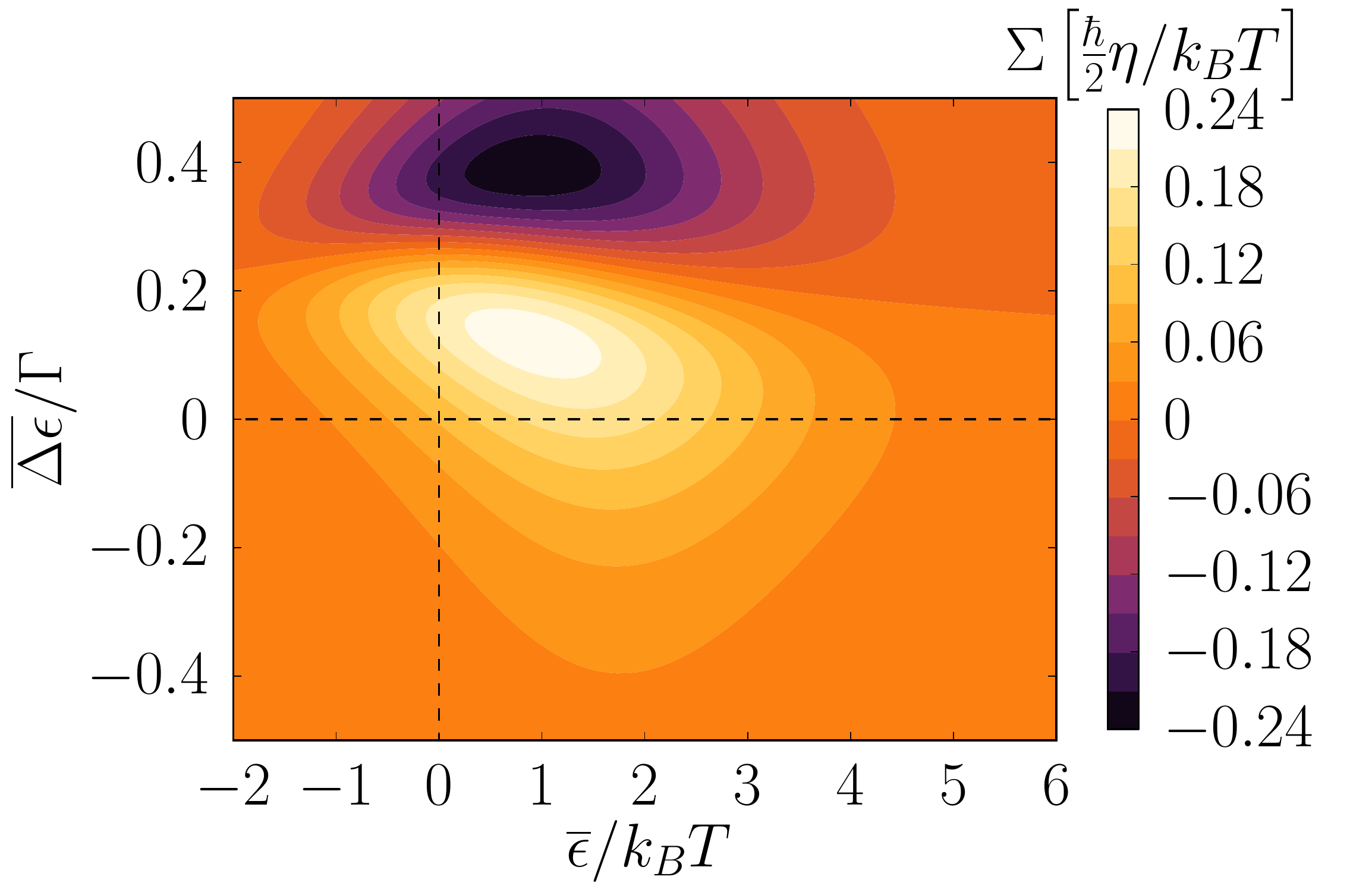}}
	\caption{\label{fig:C}(Color online) Pumped charge (spin) with finite \so~coupling, $\aso=\Gamma/10$, in the $U=0$ and the $U=\infty$ limit depending on the time-averaged orbital level positions. There are two sets of coupling parameters: First, $\phi_\text{L} = \pi/2$ and $\phi_\text{R} = \pi/4$ for panels (a)-(d), and second, the antisymmetric combination, $\phi_\text{L} = -\pi/4$ and $\phi_\text{R} = \pi/4$, for panels (e) and (f). For all panels we chose $\Gamma_\text{L} = \Gamma_\text{R}$. The latter, antisymmetric combination leads to vanishing pumped charge.}
\end{figure*}

Motivated by the previous discussion, we first focus on a tunnel-coupling configuration with $\Gamma_\text{L}=\Gamma_\text{R}$ 
but $\phi_\text{L} \neq \phi_\text{R}$, where the pumped charge and pumped spin are, in general, finite.
Both depend on the mean dot-level positions, $\be$ and $\bdep$, which is shown in Fig.~\ref{fig:C}. Those orbital energies of the quantum dot can, in principal, be adjusted by capacitively coupled gate votages. 
Figures~\ref{fig:QU0}-\ref{fig:SUI} illustrate the pumped charge (spin) for $U=0$ and $U=\infty$ and for a tunnel-coupling configuration where the coupling to the left lead is symmetric regarding the orbitals, $\phi_{\text{L}}=\pi/2$, while the coupling to the right lead is given by $\phi_{\text{R}}= \pi/4$, i.e., $V_{\text{R}1}/V_{\text{R}2}=\cot \pi/8$. 
In the case where orbital $1$ is symmetrically $\left(V_{\text{L}1}=V_{\text{R}1}\right)$ and orbital $2$ is antisymmetrically $\left(V_{\text{L}2}=-V_{\text{R}2}\right)$ coupled to the left and right leads, the pumped spin is in general finite while the pumped charge vanishes for this configuration. In Figs.~\ref{fig:PSU0} and \ref{fig:PSUI} the pumped spin is exemplarily calculated for $-\phi_\text{L}=\phi_\text{R}=\pi/4$, which is equivalent to $V_{\text{L}1}=V_{\text{R}1}=- V_{\text{L}2} \cot \pi/8=V_{\text{R}2}\cot \pi/8$.

Each plot in Fig.~\ref{fig:C} shows a maximum and a minimum value.
For no \ci, the maximum value is located at $\be=0$ (relative to the chemical potential $\mu$ of the leads). In the limit of strong \ci, the extrema positions are shifted to values of $\be$, whose order of magnitude is given by the temperature. 
The $\bdep$-position of the maximum pumped charge (spin) depends on the tunnel coupling to the leads and the \so-coupling strength. Increasing $\aso$ also increases the maximum's position with respect to $\bdep$.
Furthermore, the pumped charge is in general larger for no \ci~apart from special tunnel-coupling configurations discussed in detail in the next section. That is not surprising since the \ci~reduces the possible transport channels through the dot by suppressing occupations of the dot with more than one electron.

\subsection{Exchange-field interaction}
\begin{figure}
	\includegraphics[width=.45\textwidth]{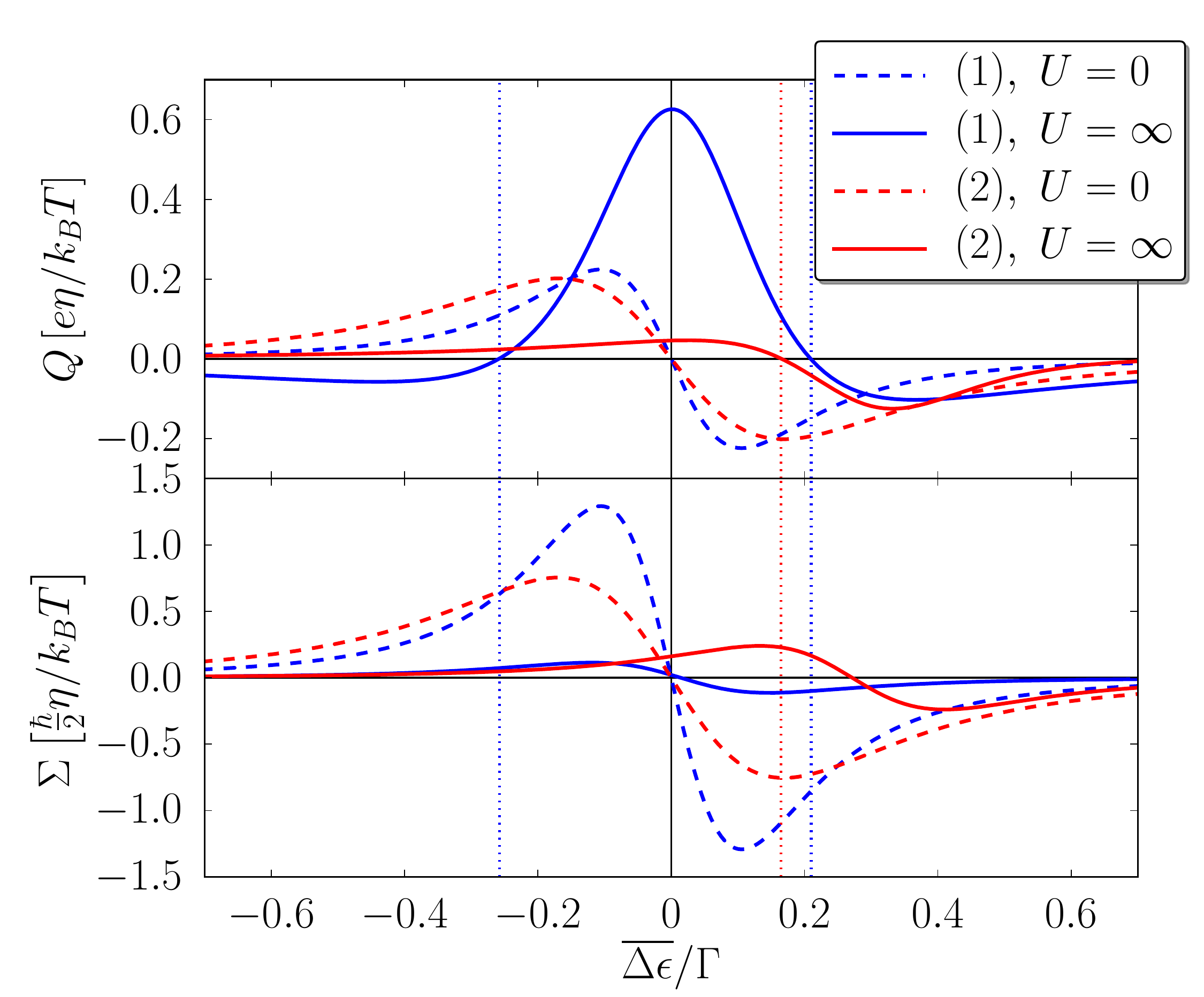}
	\caption{\label{fig:de}(Color online) Pumped charge (spin) with finite \so~coupling, $\aso=\Gamma/10$, in the $U=0$ and the $U=\infty$ limit depending on the averaged level-spacing of both orbital levels. In the $U=0$ limit we choose $\be=0$, which is the position of the maximum value. In the limit of strong \ci, we use $\be=k_BT$ as an approximation to the position of the maximum value. 
	The two sets of coupling parameters (1) and (2) are the ones given in the text (see Eq. \eqref{eq:cou}).
	The vertical dotted lines indicate \textit{pure spin pumping}. In the weak-coupling limit, this is only possible for pumping with \ci.
	}
\end{figure}

Both limits $U=0$ and $U=\infty$ show different symmetries with respect to $\bdep \rightarrow -\bdep$.  
In the limit $U=0$, the pumped charge (spin) is exactly antisymmetric in $\bdep$. 
The antisymmetry with respect to $(\be,\bdep) \rightarrow (-\be,-\bdep)$ originates from the particle-hole symmetry. 
The antisymmetry in $\bdep$ alone, on the other hand, is a non-trivial result and only valid for the lowest order contribution in $\Gamma$.
In the limit of strong \ci, the symmetry in $\bdep$ differs from the $U=0$ limit. The exchange field $B_U$, which interacts with the isospin, leads to a contribution of the pumped charge (spin) that is not antisymmetric in $\bdep$. 
Therefore, the antisymmetry is, in general, broken. 
To point out the symmetry characteristics, we study the pumped charge (spin) in two different tunnel-coupling configurations $(1)$ and $(2)$, for both $\Gamma_{\text{L}}=\Gamma_{\text{R}}$,
\begin{align}(1):\ \phi_{\text{L}}&=\frac{\pi}{4} \; ,\; \phi_{\text{R}}=\frac{2 \pi}{3}\; ,\nonumber\\
(2):\ \phi_{\text{L}}&=-\frac{\pi}{5}\; ,\; \phi_{\text{R}}=\frac{\pi}{4}\; , \label{eq:cou}\end{align}
which is equivalent to (1): $V_{\text{L}1}/V_{\text{L}2}=\cot \pi/8$, $V_{\text{R}1}/V_{\text{R}2}=1/\sqrt{3}$ in the first case, and (2): $V_{\text{L}1}/V_{\text{L}2}=-\cot \pi/10$, $V_{\text{R}1}/V_{\text{R}2}=\cot \pi/8$ in the second one.
Tunnel couplings $(1)$ and $(2)$ show that the exchange field can affect the pumped charge and the pumped spin differently, and the effect, thus, depends on the tunnel-coupling parameters. 
That is accounted for by Fig.~\ref{fig:de}, where the cut through the contour plot (of Fig.~\ref{fig:C} but with coupling configurations $(1)$ and $(2)$) for fixed $\be$ is shown. 
The fixed value of $\be$ is $\be=0$ in the limit of vanishing \ci~and, for comparison, $\be=k_BT$ in the limit of strong \ci. 

For configuration $(1)$, the exchange field leads to
a peak located at $\bdep=0$ which has a nearly symmetric behavior in $\bdep$. The pumped spin, on the other hand, is still approximately antisymmetric in $\bdep$. 
Furthermore, without $B_U$, the pumped charge (spin) is usually smaller for $U=\infty$, compared to $U=0$, because of the reduced number of transport channels through the dot, but the exchange field can enhance the pumped charge. There are sets of parameters where the charge transport is even larger for finite \ci~than for $U=0$. 

For tunnel coupling $(2)$, the symmetric part of the exchange-field contribution is less important. The pumped charge, in this case, is not dominated by a symmetric behavior as we observed for coupling $(1)$. It is, rather, a shift of the point of zero pumped charge to a finite value of $\bdep$ similar to the pumped spin.  

Comparing the exchange-field contribution for configurations $(1)$ and $(2)$, the contribution to the pumped spin reaches its maximum where the contribution to the pumped charge vanishes, and it is approximately half of its absolute maximum value where the contribution to the pumped charge has its maximum. 
For large values of exchange-field contribution, near its maximum, the pumped charge has a dominant symmetric contribution while the exchange-field contribution to the pumped spin is, in general, too small to generate a peak at $\bdep=0$.   

\subsection{Pure spin pumping\label{resPureSpin}}

Pure spin pumping is achieved whenever the pumped charge vanishes but the pumped spin remains finite.
To find such points it is helpful that the pumped charge and pumped spin behave differently in the presence of \ci, as discussed in the previous section, and that the pumped charge is more sensitive to symmetry in the tunnel-matrix elements than the pumped spin. 
This defines the two strategies to obtain pure spin pumping: to tune either the orbital energy levels of the dot or the tunnel-matrix elements.

\subsubsection{pure spin pumping by tuning orbital energies}
For fixed tunnel couplings, we try to tune the orbital energies such that the pumped charge vanishes but the pumped spin remains finite. 
As discussed above, this is easily possible for strong \ci, because in this case, the value of $\bdep$ at which the pumped charge changes its sign is shifted away from $\bdep=0$ due to the exchange field $\bm{B}_U$.
In absence of \ci~(and to lowest order in the tunnel coupling strength), this does, in general, not work apart from special coupling configurations, where the pumped charge vanishes independently of the orbital energies, as discussed in the next section.
The reason is that both the pumped charge and the pumped spin are, to lowest order in $\Gamma$, exactly antisymmetric in $\bdep$, i.e., the pumped charge and spin vanish simultaneously.
The comparison between the two limits is shown in Fig.~\ref{fig:de}.
The points of pure spin pumping are indicated by the vertical dotted lines.
Another interesting feature of the finite difference between the zero-points for the pumped charge and the pumped spin is the possibility to change the sign of the pumped spin, while charge is pumped in the same direction.

\subsubsection{pure spin pumping by tuning tunnel couplings}
\begin{figure*}
	\subfigure[~Pumped charge for $U=0$]{\includegraphics[width=.45\textwidth]{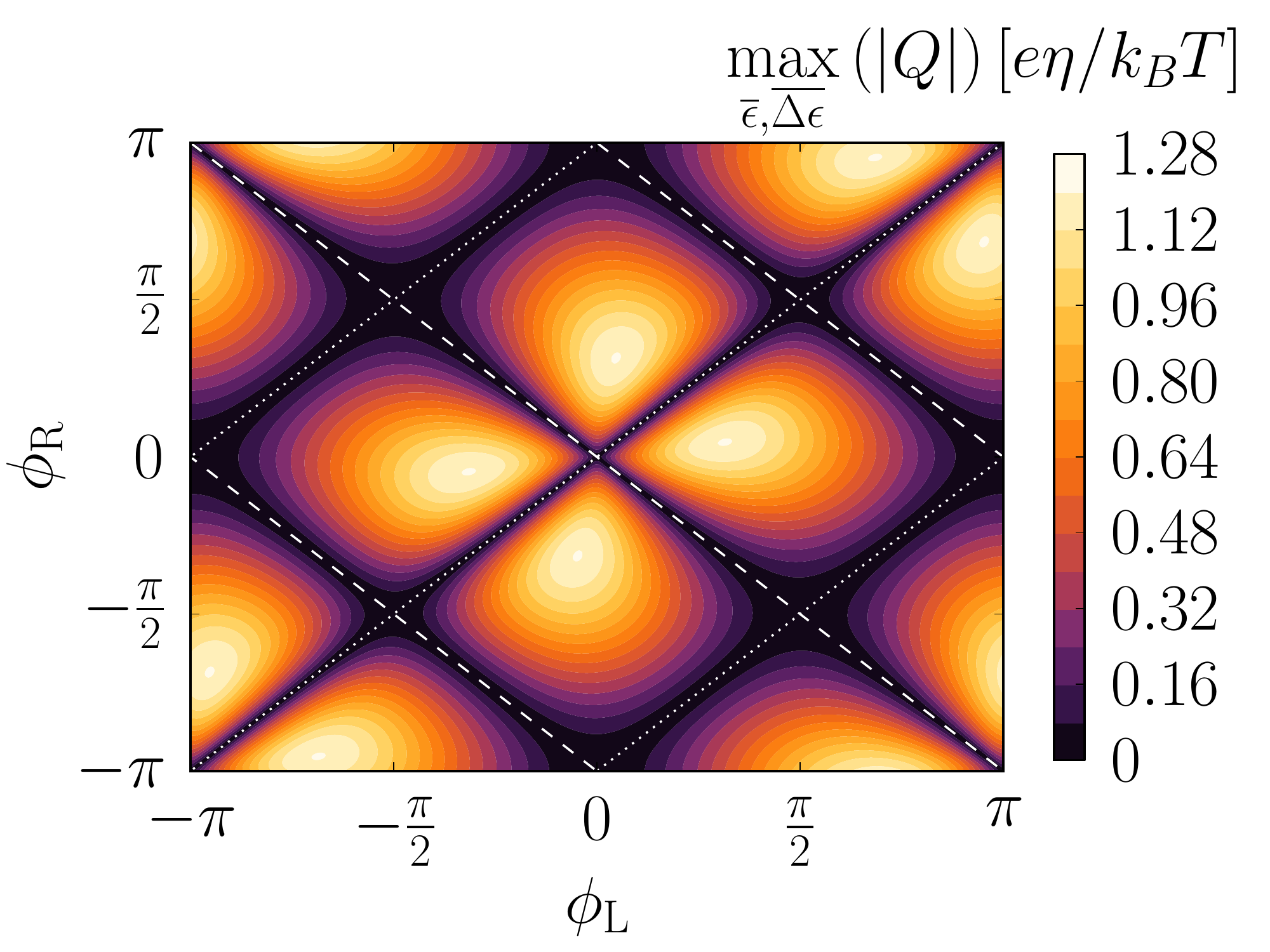}}
	\subfigure[~Pumped charge for $U=\infty$]{\includegraphics[width=.45\textwidth]{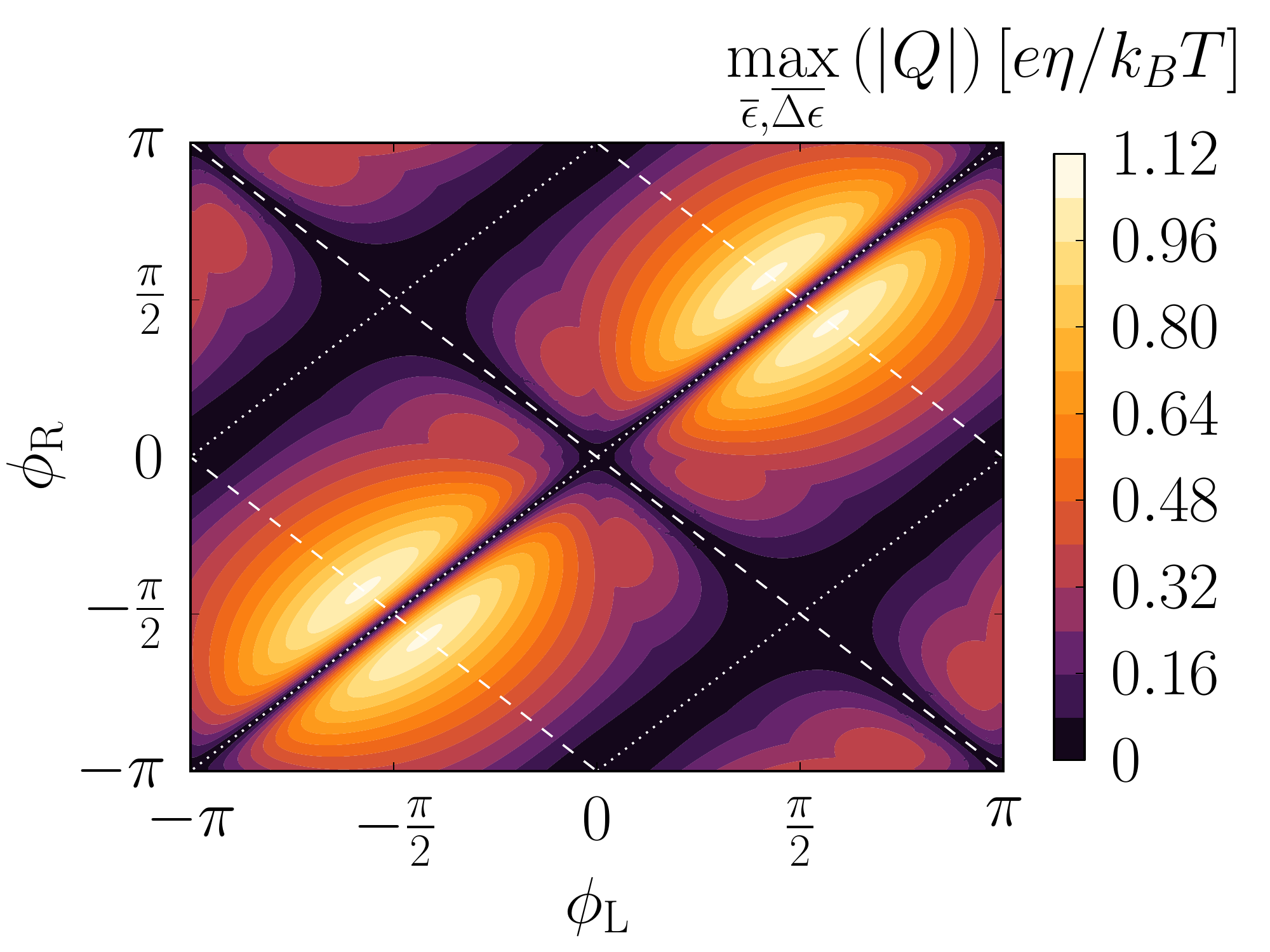}}
	\subfigure[~Pumped spin for $U=0$]{\includegraphics[width=.45\textwidth]{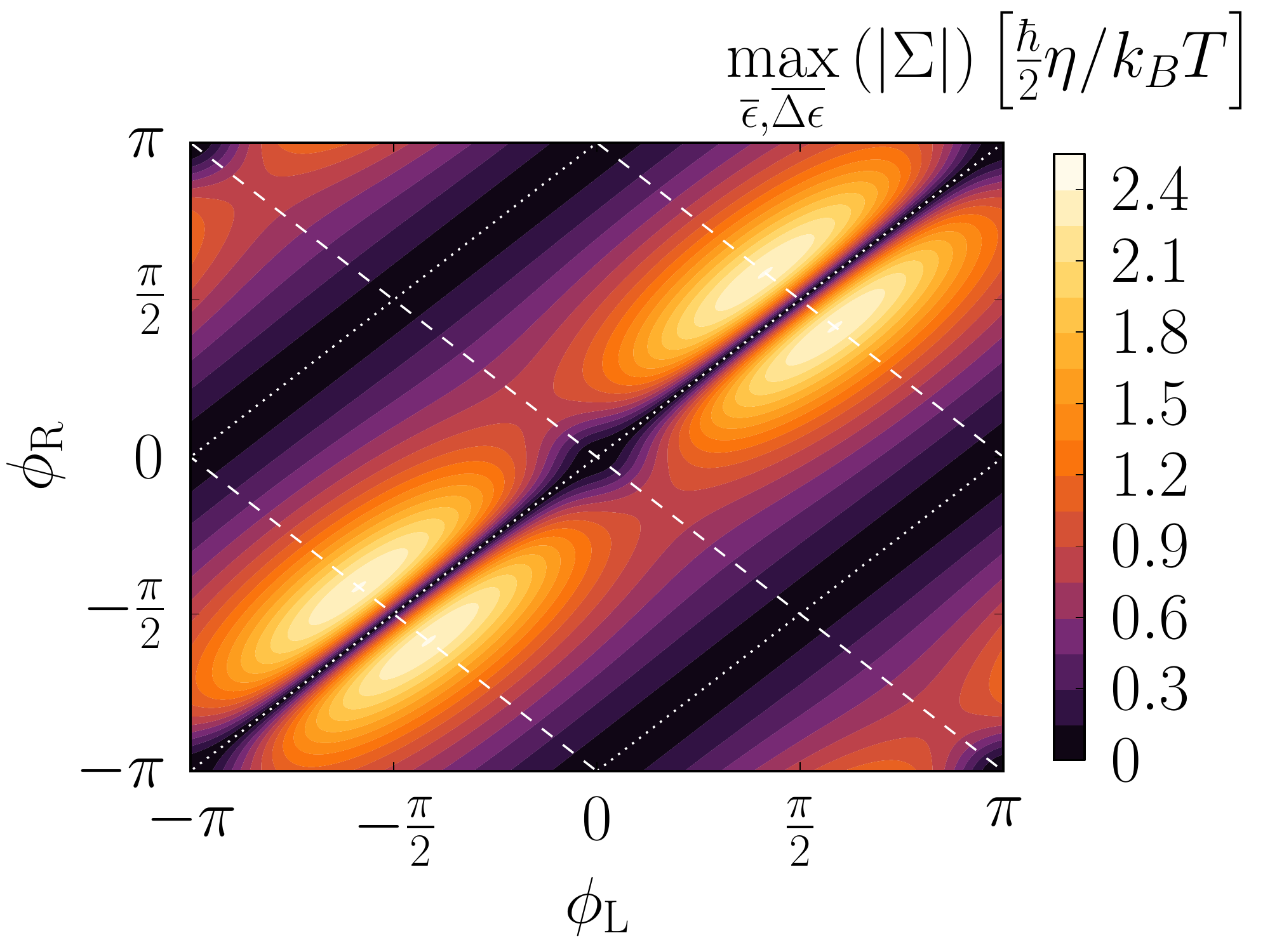}}
	\subfigure[\label{fig:CWSUI}~Pumped spin for $U=\infty$]{\includegraphics[width=.45\textwidth]{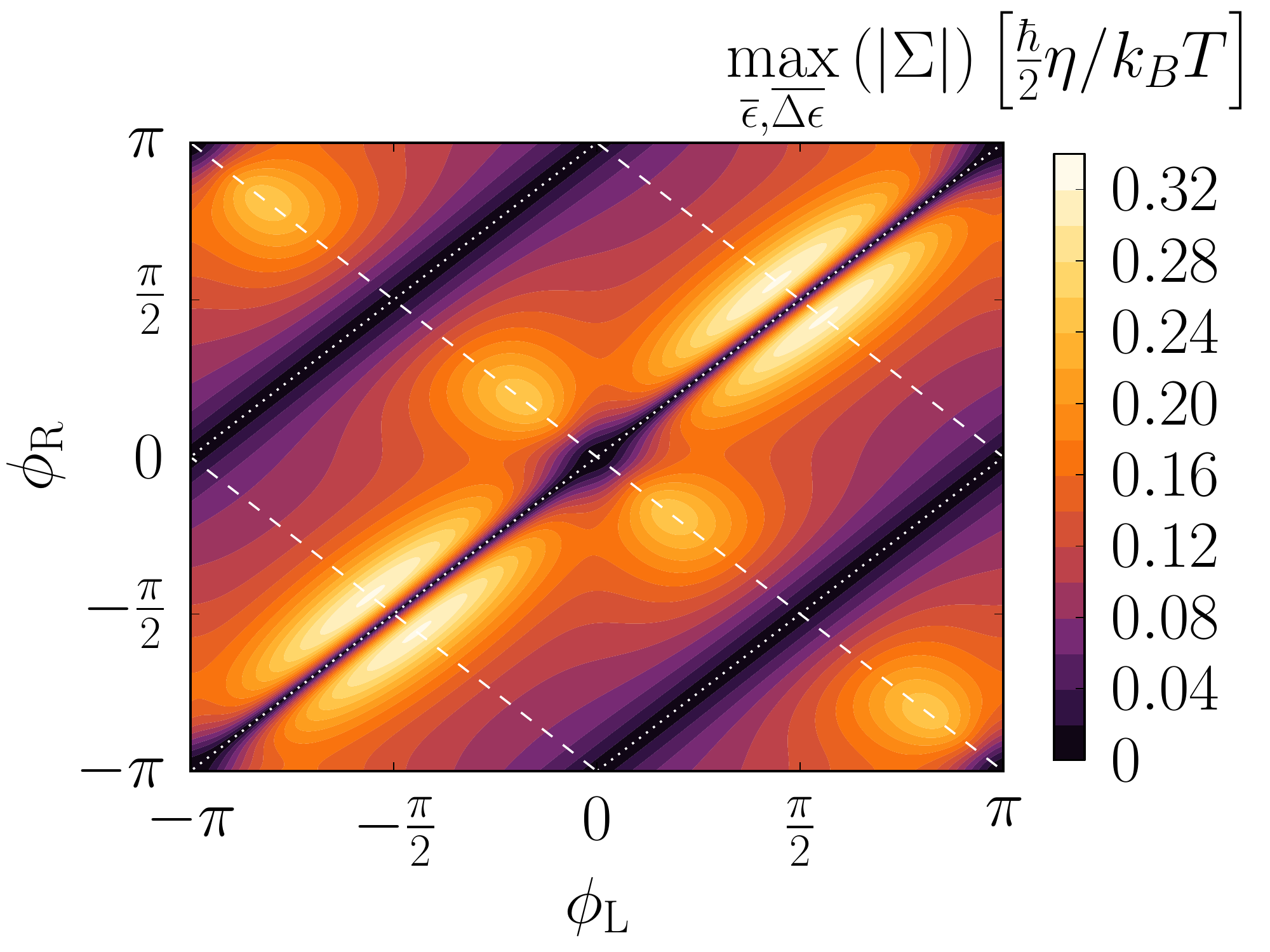}}
	\caption{\label{fig:CW}(Color online) Pumped charge and pumped spin in the $U=0$ and $U=\infty$ limits depending on the orbital-coupling configuration for fixed $\Gamma_{\text{L}}=\Gamma_{\text{R}}$. The illustrated function shows the maximum value of the pumped charge (spin) in the $(\be,\bdep)$ parameter space for $\aso=\Gamma/10$. The dotted lines represent coupling configurations where the pumped charge (spin) is zero. Along the dashed lines, for $U=0$, the pumped charge is always zero while the spin is still finite. For strong \ci, $U=\infty$, only the line $\phi_{\text{L}}=-\phi_{\text{R}}$ leads to vanishing pumped charge.}
\end{figure*}

There are cases in which pure spin current is not only possible for special, fine-tuned orbital energies but for {\em all} values of $\be$ and $\bdep$.
This is illustrated in Fig.~\ref{fig:CW}, which shows the maximum absolute value of the pumped charge (spin) in the $(\be,\bdep)$ parameter space, as a function of the coupling parameters $\phi_\la$ for $\Gamma_{\text{L}}=\Gamma_{\text{R}}$ and for both limits $U=0$ and $U=\infty$. 
The plots can be periodically continued.
The dotted lines represent coupling configurations where the pumped charge and the pumped spin are zero.
 
Along the middle dotted line, $\phi_{\text{R}}=\phi_{\text{L}}$, pumped charge and pumped spin vanish due to left-right symmetry as mentioned previously. Here, the tunnel-matrix elements are equal for the coupling to the left and the right lead, $V_{\text{L}\n}=V_{\text{R}\n}$.
The dotted zero-lines $\phi_{\text{R}}=\phi_{\text{L}}\pm\pi$ for zero pumped charge (spin) only exist for lowest order in $\Gamma$; higher-order corrections would lead, in general, to a finite pumped charge (spin). The latter conclusion can be drawn by comparing with calculations for $U=0$ which are exact in $\Gamma$, e.g., by means of a scattering matrix approach,\cite{bttiker_current_1994,brouwer_scattering_1998,avron_geometry_2000,brosco_prediction_2010} and it is self-evident that even finite \ci~does not change that significantly. Along these lines the tunnel-matrix elements are given by $V_{\text{L}1}V_{\text{R}1}=-V_{\text{L}2}V_{\text{R}2}$.
In any case, these dotted lines do no mark good candidates for pure spin pumping since, there, charge and spin behave similar.

The situation differs along the dashed lines. 
The middle dashed line, $\phi_{\text{R}}=-\phi_{\text{L}}$, represents a configuration where for each orbital the absolute value of the tunnel-matrix elements is the same, but one element of all four has an opposite sign, i.e. $V_{\text{L}1}=V_{\text{R}1}$ and $V_{\text{L}2}=-V_{\text{R}2}$ (or equivalently $1\leftrightarrow2$).
Here, we find (to lowest order in $\Gamma$) pure spin pumping for both vanishing and strong \ci.
This generalizes the result found in Ref.~\onlinecite{brosco_prediction_2010} for the $U=0$ limit to the limit of strong \ci. 
The dependence of the pure pumped spin for $\phi_{\text{R}}=-\phi_{\text{L}}=\pi/4$ on $\be$ and $\bdep$ in both Coulomb regimes is shown in Figs.~\ref{fig:PSU0} and \ref{fig:PSUI}.

The dashed lines $\phi_{\text{R}}=-\phi_{\text{L}}\pm\pi$ (equivalent to $V_{\text{L}1}V_{\text{R}1}=V_{\text{L}2}V_{\text{R}2}$) indicate a further scenario for pure spin pumping to lowest order in $\Gamma$ in the $U=0$ limit.
For higher orders in $\Gamma$, however, the pumped charge becomes finite.
It also becomes finite for $U=\infty$ (and lowest order in $\Gamma$) as a consequence of the exchange field acting on the isospin. 

\begin{figure*}
	\subfigure[~Pumped charge for $U=0$]{\includegraphics[width=.45\textwidth]{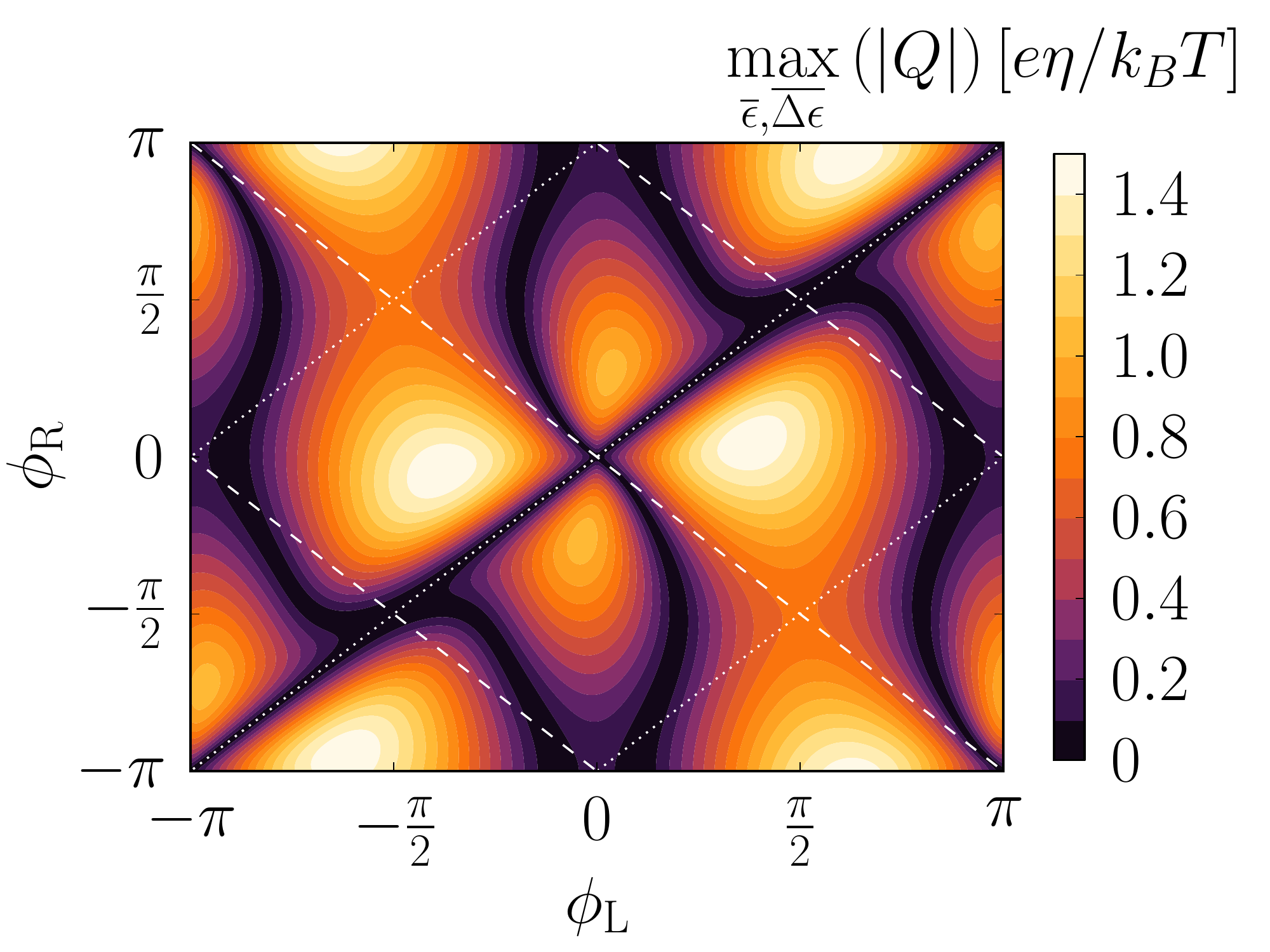}}
	\subfigure[~Pumped charge for $U=\infty$]{\includegraphics[width=.45\textwidth]{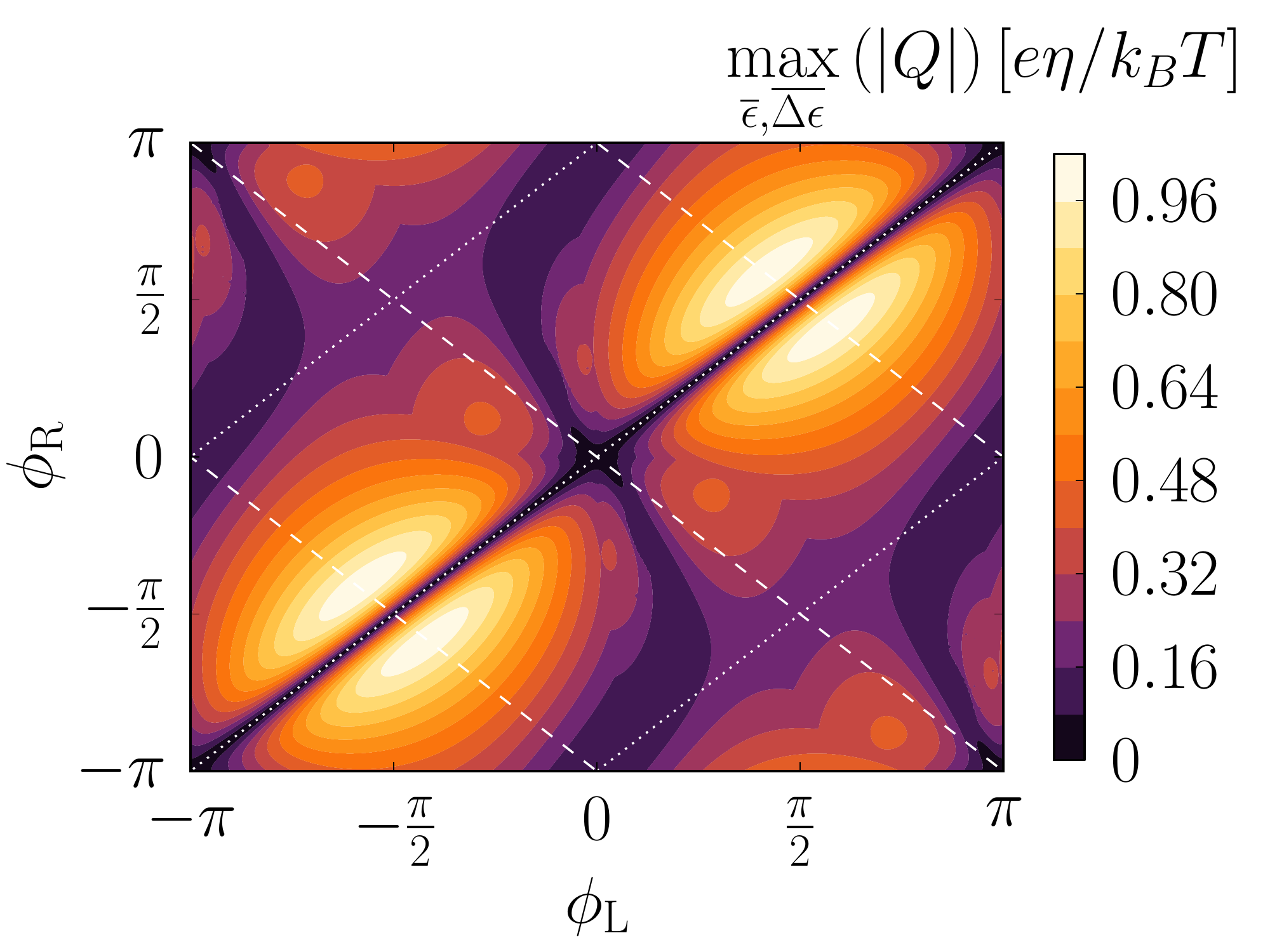}}
	\subfigure[~Pumped spin for $U=0$]{\includegraphics[width=.45\textwidth]{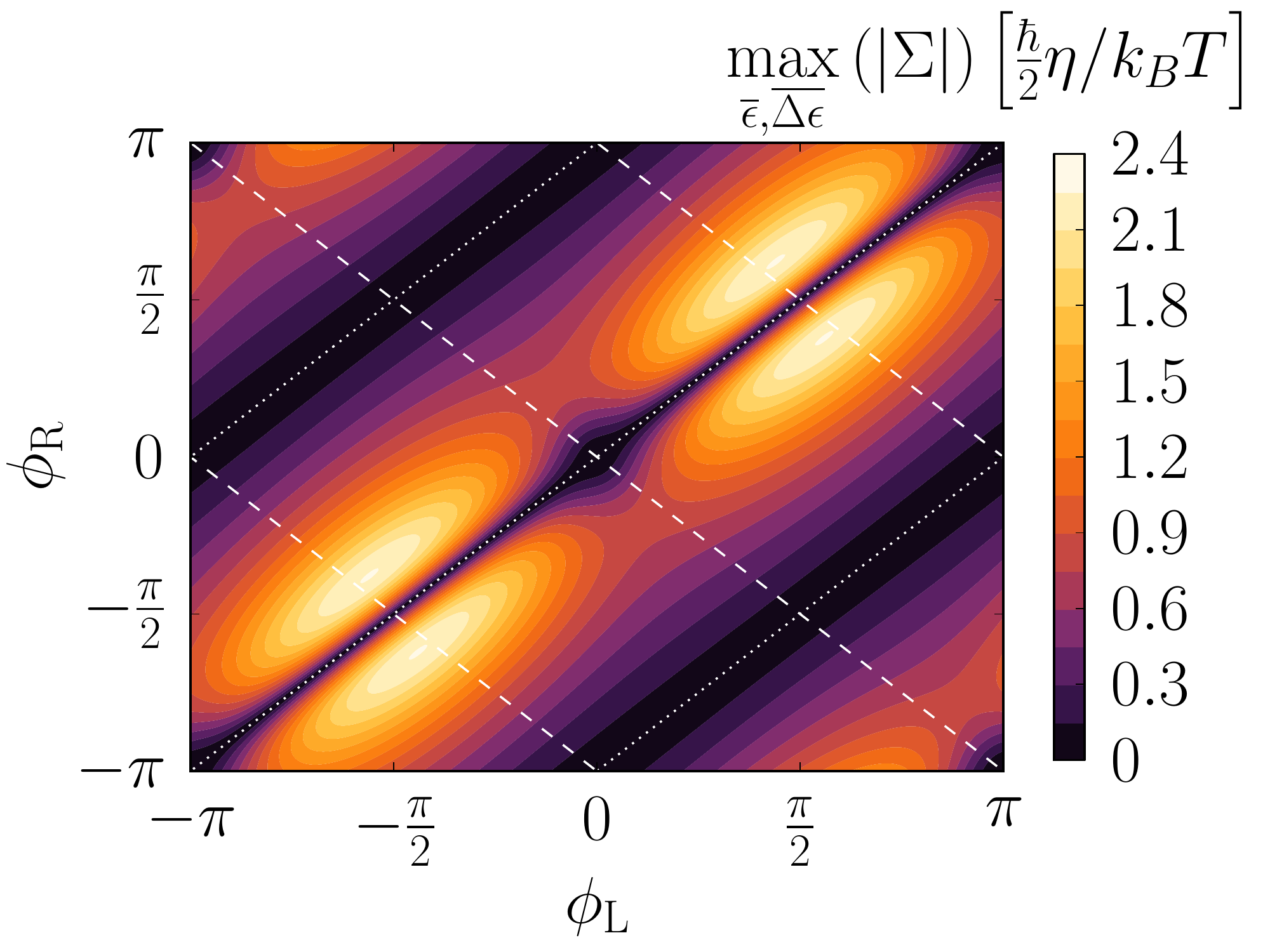}}
	\subfigure[~Pumped spin for $U=\infty$]{\includegraphics[width=.45\textwidth]{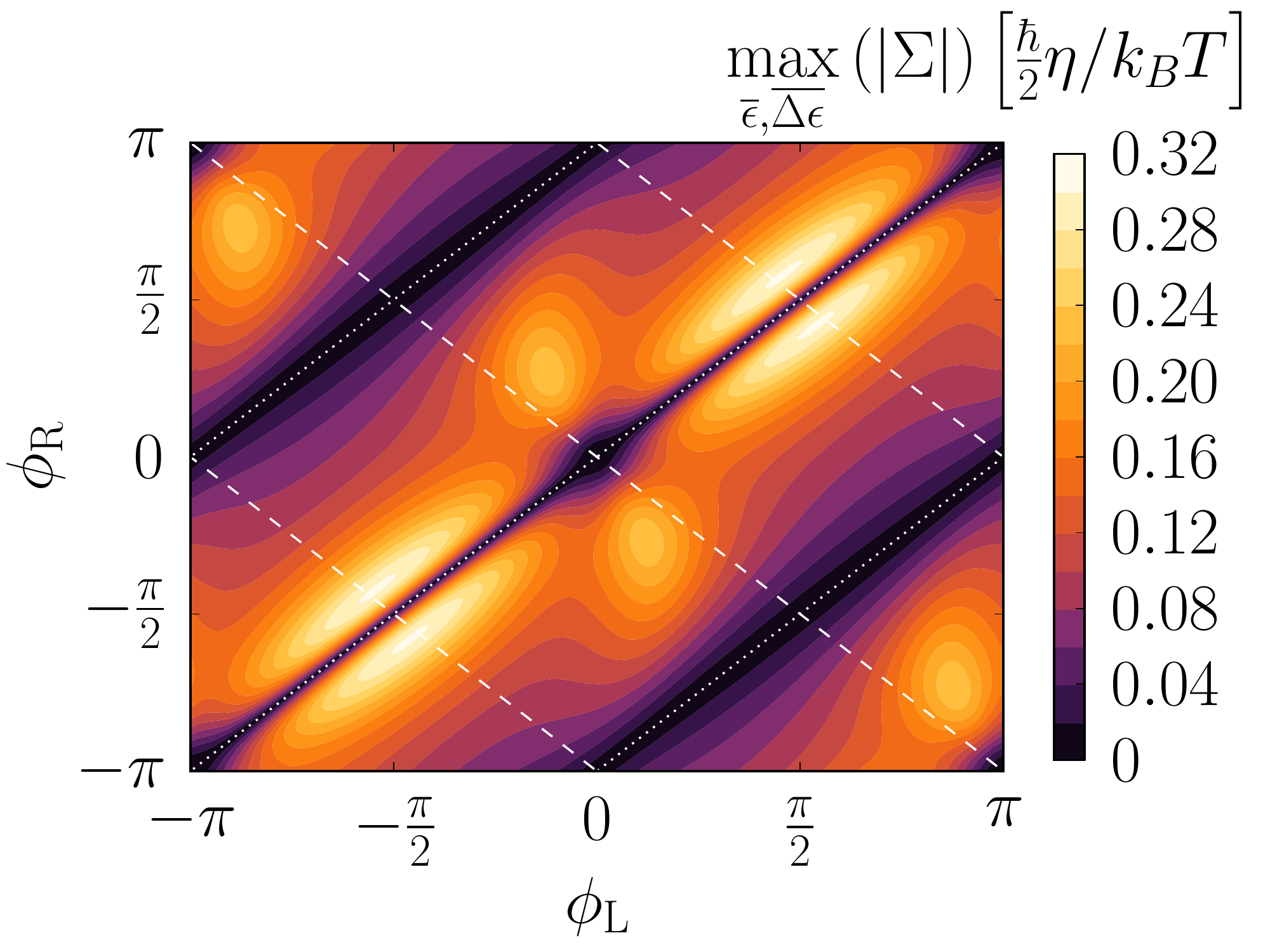}}
	\caption{\label{fig:CWa}(Color online) Pumped charge and pumped spin in the $U=0$ and $U=\infty$ limits depending on the orbital-coupling configuration for fixed $\Gamma_{\text{L}}=2 \Gamma_{\text{R}}$. The illustrated functions show the maximum value of the pumped charge (spin) in the $(\be,\bdep)$ parameter space for $\aso=\Gamma/10$. }
\end{figure*}

How important is the symmetry $\Gamma_{\text{L}}=\Gamma_{\text{R}}$?
To answer this question, we calculate the pumped charge and spin for $\Gamma_{\text{L}}=2 \Gamma_{\text{R}}$; see Fig.~\ref{fig:CWa}. 
As we see, the dependence of the pumped charge and spin on $\phi_\la$ changes substantially for the pumped charge but not so much for the pumped spin.
In particular, there are no straight lines with pure spin pumping anymore.
For $U=0$ (and to lowest order in $\Gamma$), pure spin pumping is still possible on curved lines in the $\phi_\la$ parameter space but not for $U=\infty$.
Therefore, $\Gamma_{\text{L}}= \Gamma_{\text{R}}$ is a necessary requirement for pure spin pumping.

\subsection{Spin-orbit coupling strength}
 
\begin{figure}
	\includegraphics[width=.45\textwidth]{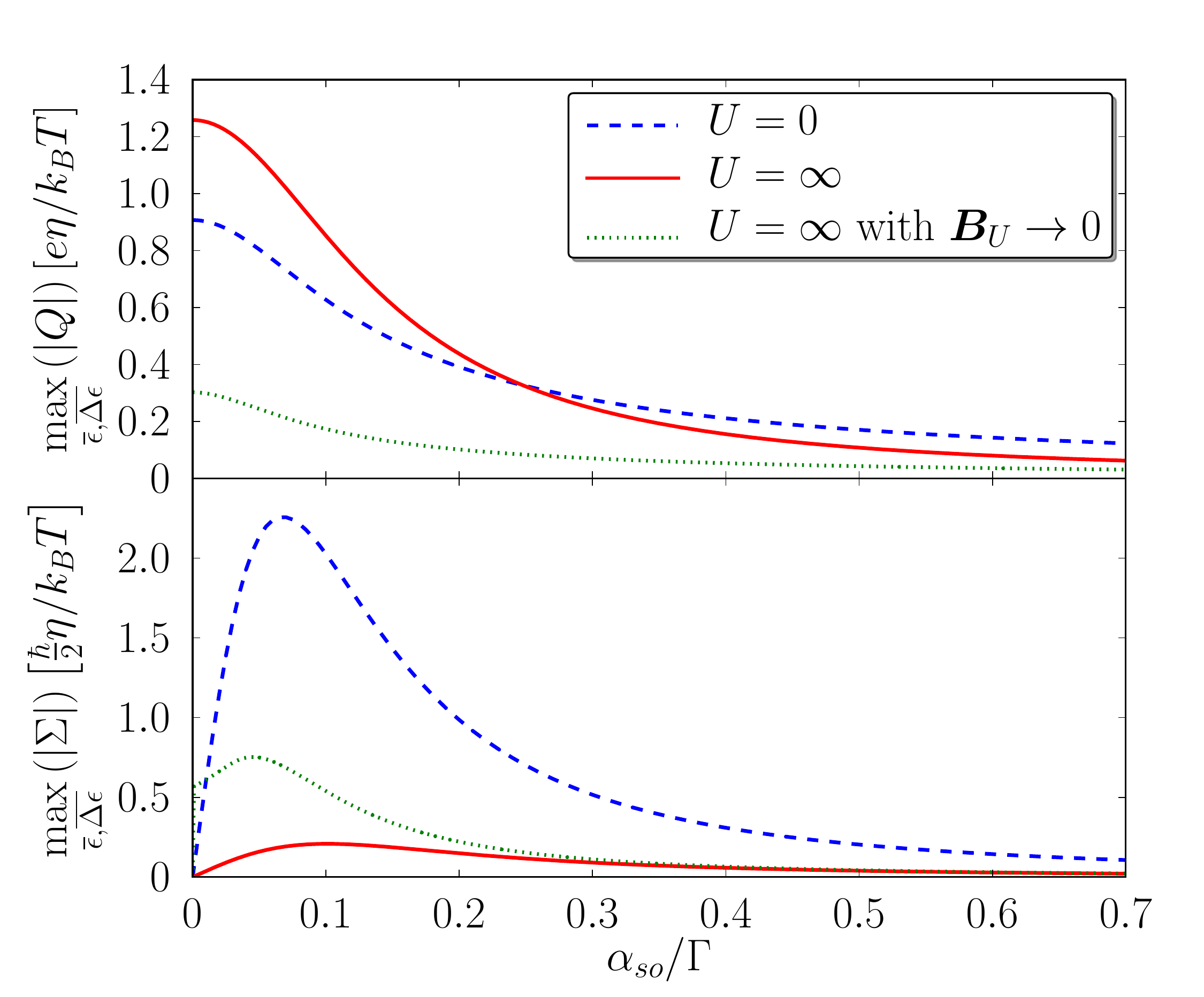}
	\caption{\label{fig:aso}(Color online) Pumped charge and pumped spin in the $U=0$ and $U=\infty$ limits depending on the strength of the \so~coupling. Additionally, the green dotted line shows the pumped charge (spin) for $U=\infty$ in the case that the exchange field is turned off by hand. The illustrated functions are the maximum value of the pumped charge (spin) for given \so~strength, $\aso$, in the $(\be,\bdep)$ parameter space. The coupling parameters are: $\Gamma_{\text{L}}=\Gamma_{\text{R}}$, $\phi_{\text{L}}=\pi/2$, and $\phi_{\text{R}}=\pi/4$. }
\end{figure}

The dependence of the pumped charge and pumped spin on the \so-coupling strength is visualized in Fig.~\ref{fig:aso}. 
Here, the different functions again show the maximum value of the absolute pumped charge (spin) in the $(\be,\bdep)$ parameter space. 
 As can be seen from the upper plot, the pumped charge decreases with increasing \so~coupling. It also decreases with increasing $\bdep$.
In both cases, the pumping is suppressed since the difference of the eigenenergies of the dot Hamiltonian becomes large.

In general, the \ci~reduces the amount of pumped charge and pumped spin. For small values of $\aso$ compared to $\Gamma$, however, the \ci~has the opposite effect on the pumped charge. In this regime, the \ci~increases the pumped charge compared to the limit of $U=0$. The latter is an effect of the exchange field: Without the exchange field, the pumped charge would be reduced due to the \ci. Increasing $\aso$ decreases the influence of the exchange field, i.e., for large $\aso$ the \ci~again reduces the pumped charge.
For the pumped spin, the situation differs: The exchange field reduces the pumped spin even further.
 
The pumped spin, in contrast to the pumped charge, vanishes for $\aso=0$. Therefore, there is an optimal value of $\aso$ that maximizes the pumped spin (see Fig.~\ref{fig:aso}). This value is smaller than $\Gamma$ and it depends on the tunnel coupling.

\section{Conclusion\label{conclusion}}
We analyze the possibility to build an all-electric spin battery and to generate a pure spin current with a two-level quantum dot in the presence of \ci.
In the limit of vanishing \ci, both are possible, as has been demonstrated in Ref.~\onlinecite{brosco_prediction_2010}.
Here, we show that this is also possible for the experimentally relevant case of a quantum dot with large \ci. 
The \ci\ changes the pumping characteristics substantially. 
In particular, symmetries with respect to the orbital energies change as a consequence of an effective exchange field acting on an isospin defined by the orbital level index. 
The nonvanishing \ci\ opens the possibility to achieve a pure spin current by tuning the orbital levels in the weak tunnel-coupling limit.
Furthermore, we find that a pure spin current is obtained independently of the orbital level energies for a certain configuration of tunnel couplings, where one level is symmetrically and the other one antisymmetrically coupled to the left and right lead, $V_{\text{L}1}=V_{\text{R}1}$ and $V_{\text{L}2}=-V_{\text{R}2}$ in terms of tunnel-matrix elements.

\acknowledgments
We acknowledge financial support from the DFG via SPP 1285 and the EU under Grant No. 238345 (GEOMDISS).

\appendix
\section{Diagrammatic rules\label{rules}}
We now specify the diagrammatic rules to calculate the diagrams of the kernels $W^{(i,n)\chi_1\chi_1^{\prime}}_{t\phantom{(,n)}\chi_2\chi_2^{\prime}}$ with $n$ tunneling lines based on Refs. \onlinecite{knig_zero-bias_1996,knig_resonant_1996,splettstoesser_adiabatic_2006,sothmann_transport_2010}. Throughout the presented calculations, only the diagrams with one tunneling line $\K^{(i,1)}_t$ are necessary.
\begin{enumerate}
	\item Draw all topologically different irreducible diagrams with $n$ tunneling lines and the dot eigenstates 
	$\chi \in  \{ \ket{0}, \ket{1}, \ket{2}, \ket{d} \}_\s$, for  $U=0$, and $\chi \in\{ \ket{0}, \ket{+\up}, \ket{-\up}, \ket{+\down}, \ket{-\down} \}$, for $U=\infty$,
	contributing to $W^{(i,n)\chi_1\chi_1^{\prime}}_{t\phantom{(,n)}\chi_2\chi_2^{\prime}}$. Each segment of the upper and lower contour separated by vertices is assigned with the corresponding eigenenergy  $E_\chi(t)$. Each tunneling line is labeled with the lead $\la$, spin $\s$ and energy $\w$.
	\item Each time segment of the diagram between two vertices at the times $t_j$ and $t_{j+1}$ leads to a contribution $1/(\de_j+i 0^+)$, where $\de_j$ is the difference of left going energies minus right going energies.
	\item Each tunneling line that goes forward or backward with respect to the Keldysh contour contributes with a factor $(1-f(\w))$ or $f(\w)$, respectively, where $f(\w)$ is the Fermi function. Furthermore, a tunneling line that begins at a vertex containing a dot operator $\dab{\gamma\s}$, with $\gamma=\pm$, and ends at a vertex containing $\dauf{\gamma\p\s}$ introduces a factor $\tilde{\Gamma}_{\s\la\gamma\p\gamma}/2 \pi$. The matrix elements $\tilde{\Gamma}_{\s\la\gamma\p\gamma}$ are obtained from the transformation $\bm{\tilde{\Gamma}}_{\s\la}=\bm{T}^\dagger_\s \bm{\Gamma}_\la \bm{T}_\s$, with $\Gamma_{\la \n \n\p}=2\pi\rho V_{\la\n\p}V^\ast_{\la\n}$, where $\n,\n\p=1,2$.
	\item Each vertex in the $U=0$ limit that connects state $\ket{-}$ with state $\ket{d}$ gives rise to a minus sign.  
	\item The overall prefactor is $-\frac{i}{\hbar}(-1)^{b+c}$, where $b$ is the number of vertices on the lower contour line and $c$ the number of crossings in the tunneling lines. 
	\item Integrate over all energies of the tunneling lines and sum over  $\la$ and $\s$. Sum up all contributing diagrams.
\end{enumerate}

\end{document}